%% file: main.tex
\documentclass[12pt]{article}
\usepackage[utf8]{inputenc}
\usepackage[T1]{fontenc}
\usepackage{palatino}
\usepackage[a4paper, left=2cm, right=2cm, top=2cm, bottom=2cm]{geometry}
\usepackage{algorithm}
\usepackage{lettrine}
\usepackage{amssymb}
\usepackage{amsmath} 
\usepackage{amsthm}
\usepackage{mathabx}
\usepackage{dsfont}
\usepackage[square,numbers]{natbib}
\usepackage{hyperref}
\usepackage{graphicx}
\usepackage{subcaption}
\usepackage{import}
\usepackage{float}

\DeclareMathOperator*{\var}{Var}
\DeclareMathOperator*{\cov}{Cov}

\newtheorem{lemma}{Lemma}
\import{./}{commands.tex}

\usepackage{colortbl}

\newcommand{\cy}[1]{{\color{black}{#1}}}

\title{Uncertainty quantification and global sensitivity analysis of seismic fragility curves using kriging}
\author{Clément Gauchy$^{1, 2}$, Cyril Feau$^{1}$, Josselin Garnier$^{2}$}
\date{$^1$ \small{Université Paris-Saclay, CEA, Service d'\'Etudes Mécaniques et Thermiques, 91191, Gif-sur-Yvette, France} \\%
    $^2$ \small{CMAP, \'Ecole Polytechnique, Institut Polytechnique de Paris, 91128 Palaiseau Cedex, France} \\[2ex]%
    }

\begin{document}
\maketitle

\begin{abstract}
   Seismic fragility curves have been introduced as key components of Seismic Probabilistic Risk Assessment studies. They express the probability of failure of mechanical structures conditional to a seismic intensity measure and must take into account the inherent uncertainties in such studies, the so-called epistemic uncertainties (i.e. coming from the uncertainty on the mechanical parameters of the structure) and the aleatory uncertainties (i.e. coming from the randomness of the seismic ground motions). For simulation-based approaches we propose a methodology to build and calibrate a Gaussian process surrogate model to estimate a family of non-parametric seismic fragility curves for a mechanical structure by propagating both the surrogate model uncertainty and the epistemic ones. Gaussian processes have indeed the main advantage to propose both a predictor and an assessment of the uncertainty of its predictions. In addition, we extend this methodology to sensitivity analysis. Global sensitivity indices such as aggregated Sobol indices and kernel-based indices are proposed to know how the uncertainty on the seismic fragility curves is apportioned according to each uncertain mechanical parameter. This comprehensive Uncertainty Quantification framework is finally applied to an industrial test case consisting in a part of a piping system of a  Pressurized Water Reactor.
\end{abstract}

    \section{Introduction}
 	\cy{In the 1980s, a probabilistic framework was developed} to evaluate the \cy{mean} annual probability of occurrence of severe damage on structures caused by seismic ground motions, coined Seismic Probabilistic Risk Assessment (SPRA) \cite{Kennedy1980, Kennedy1984, BakerCornell2008}. \cy{One of the key elements of this approach is the fragility curve. Such a curve expresses the probability of failure (or undesirable outcome) of a structure conditional to a seismic intensity measure and must take into account the different sources of uncertainties that inevitably come into play in this type of study and which are classified into two categories, namely:  the \textit{epistemic} and the \textit{aleatory} uncertainties. According to \cite{DerKiureghian2009}, distinguishing between these two types of uncertainties is a pragmatic way of distinguishing which uncertainties engineers can reduce and which cannot, allowing for  information based design choices. For that reason, in practice, it is often assumed that epistemic uncertainties are sources of uncertainty that can be reduced in the short term with a reasonable budget, while aleatory uncertainties are devolved to sources of natural hazards due to physical phenomena. Thus, 
 	%in the spirit of the pioneering work of the 1980s \cite{Kennedy1980, Kennedy1984}, 
 	a seismic fragility curve is not strictly speaking a single curve (i.e. mean curve), but a family of fragility curves which reflects the uncertainty on the mean seismic fragility curve due to a certain lack of knowledge of the structure of interest and its environment (i.e including soil–structure interaction, etc.).}
	
	\cy{Since the 1980s several techniques have been developed to estimate such curves, in the sense of mean fragility curves most of the time. When little data is available, whether experimental, from post-earthquake feedback or from numerical calculations, a classic approach to circumvent estimation difficulties is to use a parametric model of the fragility curve, such as the lognormal model historically introduced in \cite{Kennedy1980} (see e.g. \cite{Wang2020, Shinozuka00, ELLINGWOOD2001251, Lallemant2015, Mai2017, Trevlopoulos2019}). As the validity of parametric models is questionable, non-parametric estimation techniques have also been developed, such as kernel smoothing \cite{Lallemant2015, Mai2017} as well as other methodologies \cite{Trevlopoulos2019, ALTIERI2020}. Most of these strategies are compared in \cite{Lallemant2015, Mai2017, Baker2015} and \cite{Lallemant2015} presents their advantages and disadvantages. Beyond these methods, techniques based on statistical and machine learning on the mechanical response of the structure can also be used, including: linear or generalized linear regression \cite{Lallemant2015}, classification - based techniques \cite{Bernier2019, Sainct2020}, polynomial chaos expansion \cite{Mai2016, Zhu2020} and artificial neural networks \cite{WANGZ2018, Sainct2020}. Most of these techniques take advantage of the rise of computational power %since the end of the 1990s 
	to allow estimations based on numerical simulations. They make it possible to reduce the computational burden which remains high because such estimations require a large number of numerical simulations to be precise. Nevertheless, despite all these techniques, one of the main challenges that persists is the estimation at a lower numerical cost (i.e. with few calls for computer codes) of non-parametric fragility curves taking into account the two types of uncertainties.}
	%in the logic of the pioneering work of the 1980s.}

    \cy{The objective of this work is to propose a methodology that meets these requirements in a numerical simulation based framework. As we focus on approaches based on numerical simulations that rely on real seismic signal databases enriched by means of a seismic signal generator that well encompasses their temporal and spectral non-stationarities \cite{Rezaeian10}, we assume that there is no epistemic uncertainty affecting the excitation which only represents the aleatory uncertainty of the problem. Consequently, in our settings, epistemic uncertainties only concern the mechanical parameters of the structures of interest. The physics-based approaches developed as part of Performance-Based Earthquake Engineering (PBEE) address this problem \cite{Gardoni2002}. However, they are not suitable when the use of detailed finite element simulations is required, in order to take into account all the specificities of the structures of interest as it can be the case nowadays for the seismic safety studies in nuclear industry \cite{Wang2020, Zentner2010, Mandal2016}. So, in this paper, our approach relies on the use of surrogate models} of the computer codes, also referenced as metamodels, based on Gaussian process regression. This framework corresponds to a data driven approximation of the input/output relationship of a numerical computer code based on a set of experiments (e.g. computer model calls) at different values of the input parameters with a Gaussian process assumption on the numerical computer code output values \cite{Sacks1989}. Gaussian process regression, or kriging in the field of geostatistics, has gained in popularity because of its predictive capabilities and its ability to quantify the surrogate model uncertainty \cite{Rasmussen2006}. Gaussian process surrogates have already been used for various applications in engineering, such as seismic risk assessment \cite{GidarisTaflanidis2015, KypriotiTaflanidis2021}, thermohydraulics for safety studies of nuclear power plants \cite{MarrelIoossChabridon2021} or hydrogeology for radionucleide transport in groundwater \cite{Marrel2008}. \cy{In this work, we propose a methodology to build and calibrate a Gaussian process surrogate model to estimate a family of seismic fragility curves for mechanical structures - defined here as seismic fragility quantile curves -  by propagating both the surrogate model uncertainty and the epistemic ones.}
   
    \cy{In such a context, the use of Sensitivity Analysis (SA) techniques is essential for engineers.} Indeed, according to \cite{Saltelli2004}, SA goal is to investigate how the uncertainty of the model output can be apportioned to different sources of uncertainties of the model input. SA techniques are also performed according to a range of conceptual objectives, coined as SA settings, defined in \cite{Saltelli2004, Borgonovo2017}. These objectives are prioritizing the most influential inputs, thus a possible reduction of uncertainty affecting these inputs may lead to the largest reduction of the output uncertainty, and identifying the noninfluential inputs which then could be fixed at a given value without any loss of information about the model output. SA techniques are classically applied on the model output, however it is possible to extend \cy{their fields of application} to goal-oriented \cy{quantities} of interest such as seismic fragility curves. 
    %Moreover, the clear distinction between aleatory and epistemic uncertainties in the SPRA setting motivates the use of SA techniques for the epistemic uncertainties, which concern the mechanical parameters of the structure. 
    \cy{In our case,} SA techniques will help \cy{to determine} which mechanical parameter \cy{uncertainties most influence} the seismic fragility curve uncertainty. \cy{Note that} SA on the mechanical parameters of the \cy{structures} is peculiarly challenging, due to the strong influence of the seismic ground motions on \cy{their responses}. However, even if the uncertainty coming from mechanical parameters is smaller that the one coming from the seismic ground motion, SA on these parameters is crucial to propose information-based choices to engineers and to discuss quantitatively the different possible designs of the mechanical structure studied, especially in the context of nuclear industry where safety constraints imposed by regulatory agencies are very high. \cy{In \cite{BORGONOVO201366} CDF-based importance measures are used to address the problem of ranking of uncertain model parameters in seismic fragility analysis. To go further, we propose to use} Global Sensitivity Analysis (GSA) methods \cite{Iooss2015, DaVeiga2021} which take into account the overall uncertainty ranges of the parameters. \cy{We present} global sensitivity indices applied in the particular context of seismic fragility curves as a quantity of interest. \cy{We are first interested in} the estimation of the Sobol indices \cite{Sobol1993, Sobol2001} adapted to seismic fragility curves. We also focus on recently studied global sensitivity indices based on kernel methods \cite{Rabitz2022}, the $\beta^{k}$-indices, which seem adapted to functional quantities of interest like fragility curves. \cy{However, because the estimation of global sensitivity indices requires a large number of simulations that is intractable using complex numerical simulations, the Gaussian process surrogate is also used to estimate the global sensitivity indices on the seismic fragility curves}. Moreover, as in \cite{LeGratiet2014}, the Gaussian process surrogate uncertainty will be propagated into the global sensitivity indices estimates.
    
    \cy{This paper, which presents a comprehensive Uncertainty Quantification (UQ) framework for seismic fragility curves of mechanical structures, taking into account metamodel and mechanical parameter uncertainties, is organized as follows:} Section \ref{sec: estim frag} is devoted to the estimation of seismic fragility curves using Gaussian process regression, \cy{Section} \ref{sec: GSA method} concerns the definition of global sensitivity indices tailored for seismic fragility curves, the aggregated Sobol indices and the $\beta^k$ indices. Section \ref{sec: numerical appli} presents an illustration of the methodology developed in this article to an industrial test case consisting in a mock-up of a piping system of a French Pressurized Water Reactor (PWR).

    \section{Estimation of seismic fragility curves using Gaussian Process surrogates}\label{sec: estim frag}
    
     \cy{As discussed in the introduction, the sources of uncertainties are in this work divided into two categories, the aleatory and epistemic uncertainties. 
     
     Aleatory uncertainties are related to the stochastic ground motions. To account for them, we use a synthetic generator of ground motions to enrich a set of real seismic signals selected in a database for a given magnitude (M) - source-to-site distance (R) scenario. This generator is based on a filtered modulated white-noise process \cite{Rezaeian10}. It is common in SPRA studies to sum up the seismic hazard by a so-called Intensity Measure (IM), which is the variable against which the fragility curves are conditioned. This is often a scalar value obtained from the seismic signals such as the Peak Ground Acceleration (PGA) or the Pseudo Spectral Acceleration (PSA). In \cite{Cornell2004}, the author recalls the main assumptions according to which it is possible to reduce the seismic hazard to the IM values (see also \cite{Grigoriu2021}). In the following, we denote by $\im$ the scalar value corresponding to the IM.
     
     Epistemic uncertainties are related to the mechanical properties of the model of the structure.} These parameters are denoted by the vector $\xe \in \Xset \subset \mathbb{R}^\epidim$. Furthermore,  we denote by $\edp$ the Engineering Demand Parameter (EDP) of interest, which can be the peak inter story drift for a multistoreys building or a rotation angle \cy{of a specific elbow} of a piping system of a nuclear power plant. A very common statistical model between the EDP and the combination of structural and seismic uncertainty is the log-normal model:
    \begin{equation}\label{eq: stat model}
        \log(\edp(\im, \xe)) = \regr(\im, \xe) + \varepsilon(\im, \xe) \ ,
    \end{equation}
    where  $\xe$ is the vector of the mechanical properties of the structure, $\im$ is the IM, $\regr(a,\x)$ is the regression function, and $\varepsilon \sim \Norm(0, \sigeps(\im, \xe)^2)$ is a centered Gaussian noise. Note that this log-normal assumption for the EDP distribution is not necessary for the proposed methodology, any functional transformation of $\edp$ (such as Box-Cox transformation \cite{Box1964}) is possible as long as it is normally distributed after this transformation. For the sake of notation simplicity, we denote $\y(\im, \xe) = \log(\edp(\im, \xe))$. The fragility curve is then defined by:
    \begin{equation}\label{eq: frag}
        \fragepi(\im, \xe) = \prob(\edp(\Itm, \Xe) > C | \Itm = \im, \Xe = \xe)  ,
    \end{equation}
    where $\Itm$ is the real-valued random variable of the seismic intensity measure and $\Xe$ the random vector of the mechanical parameters of the structure. $C$ corresponds to a deterministic threshold of acceptable robustness of the structure. Substituting the model Equation \eqref{eq: stat model} into Equation \eqref{eq: frag} we get the form of the fragility curve
    \begin{equation}
          \fragepi(\im, \xe)= \Phi\left(\frac{\regr(\im, \xe) - \log(C)}{\sigeps(\im, \xe)}\right) \ , 
    \end{equation}
    where $\Phi$ is the cumulative distribution function (cdf) of the standard Gaussian distribution.
    In this framework the numerical simulations of the structure are made by a computer model. The computer model is considered of high-fidelity with respect to the mechanical problem studied and therefore it may involve a chain of multi-physics simulation codes (involving finite elements or finite volumes, computational fluid dynamics...) and thus it is considered as a \textit{black-box}. This means that the different strategies described throughout this paper \cy{are} non-intrusive with respect to this black-box computer model.
    
    \subsection{Gaussian process surrogate with homoskedastic nugget noise}
    
    In this section, we suppose that the regression function $\regr$ is a realization of a Gaussian process $G$ and the Gaussian noise $\varepsilon(\im, \xe)$ is homoskedastic and will be denoted by $\varepsilon$ such that $\varepsilon \sim \Norm(0, \sigeps^2)$. We thus define the random observation by:
    \begin{equation}\label{eq: stat model gp}
        \Ygp(\im, \xe) = \regrgp(\im, \xe) + \varepsilon \ .
    \end{equation}
    
    Remark in Equation \eqref{eq: stat model gp} that thanks to the Gaussian noise assumption on the noise $\varepsilon$, the random observations $\Ygp(\im, \xe)$ is also a Gaussian process. We make the assumption that $G$ is a zero mean Gaussian process with a tensorized anisotropic stationary Mat\'ern $5/2$ covariance function parametrized by its intensity $\sigma$ and its lengthscales $(\lengthscale_i)_{1 \leq i \leq d+1}$. This covariance function is motivated by is popularity in the machine learning community as it covers a large number of applications. Note also that with such a covariance function the Gaussian process $G$ is two times mean-square differentiable, which is a good compromise between the regularity of the regression function $g$ and the potential sparsity of the data. 
    
    Given an experimental design made of $n$ simulations of the mechanical computer model, we obtain the dataset $\dataset_n = ((\im_i, \xe_i), \y(\im_i, \xe_i))_{1 \leq i \leq n}$.  By the maximum likelihood method, we can provide estimates for the unknown covariance function hyperparameters $\sigma, (\lengthscale_i)_{1 \leq i \leq d+1}$ and also the Gaussian noise variance $\sigeps$ (see \cite{Marrel2008} for a practical implementation of the method).
    The dataset $\dataset_n$ can then be used to derive the conditional distribution of the Gaussian process $\Ygp$ for any $(\im, \xe)$:
    \begin{equation}\label{eq: conditional proba Ygp}
        (\Ygp(\im, \xe)| \dataset_n) \sim \Norm\left(\gppred_n(\im, \xe), \siggp_n(\im, \xe)^2\right) \ ,
    \end{equation}
    where $\gppred_n(\im, \xe)$ and $\siggp_n(\im, \xe)^2$ are obtained from the kriging equations \cite[p.16 - 17]{Rasmussen2006}.
    In the same fashion, we can derive the conditional distribution of the Gaussian process $\regrgp$ on the regression function for any $(\im, \xe)$:  
    \begin{equation}\label{eq: conditional proba G}
        (\regrgp(\im, \xe)| \dataset_n) \sim \Norm\left(\gppred_n(\im, \xe), \siggplatent_n(\im, \xe)^2\right) \ ,
    \end{equation}
    where $\siggp_n(\im, \xe)^2 = \siggplatent_n(\im, \xe)^2 + \sigeps^2$. The fragility curve is then obtained by replacing the computer model output $\y$ by a Gaussian process $\Ygp_n$ which follows the distribution of the Gaussian process $Y$ conditioned to $\dataset_n$ detailed in Equation \eqref{eq: conditional proba Ygp}. Hence for any vector $(\im, \xe)$ we derive the estimator of the fragility curve $\fragepigp$:
    \begin{equation}\label{eq: fragepigp def}
       \fragepigp(\im, \xe) = \prob(\Ygp_n(\im, \xe) > \log(C) | \Itm = \im, \Xe = \xe) \ .
    \end{equation}
    
    We can then use the distribution of $\Ygp_n$ to estimate the fragility curve:
    \begin{equation}\label{eq: fragepigp estimation}
        \fragepigp(\im, \xe) = \Phi\left(\frac{\gppred_n(\im, \xe) - \log(C)}{\siggp_n(\im, \xe)}\right) \ .
    \end{equation}
%    where $\Phi$ is the cdf of the standard Gaussian distribution.
    Moreover, the Gaussian process surrogate allows us to propagate the surrogate model uncertainty into the fragility curve, thanks to the conditional distribution of the regression function $(\regrgp(\im, \xe)| \dataset_n)$. We introduce $\regrgp_n$ a Gaussian process with the same distribution as the Gaussian process $(\regrgp | \dataset_n)$, then the fragility curve tainted by the uncertainty of the Gaussian process surrogate writes:
    \begin{align}
    \label{eq: fragepigp uncertainty} 
        \fragepireggp(\im, \xe) &= \Phi\left(\frac{\regrgp_n(\im, \xe) - \log(C)}{\sigeps}\right) \ ,
 \end{align}
where $\regrgp_n(\im, \xe) \sim \Norm\left(\gppred_n(\im, \xe), \siggplatent_n(\im, \xe)^2\right)$.    Remark that $\fragepigp$ is the mean of $\fragepireggp$ with respect to the distribution of $G_n$. In order to estimate the distribution of $\fragepireggp$, we simulate $P$ realizations $(\regrgp_{n,p}(\im, \xe))_{1 \leq p \leq P}$ with the distribution of $(\regrgp(\im, \xe)| \dataset_n)$ to estimate a sample of $\fragepireggp$:
    
    \begin{equation}\label{eq: fragepigp uncertainty sample}
        \fragepireggp_p(\im, \xe) = \Phi\left(\frac{\regrgp_{n,p}(\im, \xe) - \log(C)}{\sigeps}\right) \ .
    \end{equation}
    
    However, some mechanical structures have nonlinear behavior that can influence the local variability of the log-EDP $y(\im, \xe)$. Thus, a varying nugget with respect to $(a, \xe)$ is necessary to capture the form of $y(\im, \xe)$. This comes with a cost in terms of dataset size, due to the increase in the numbers of parameters to estimate. \cy{We deal with this case in the next section.} 
    %We will then discuss in which situation it is more impactful to consider a heteroskedastic Gaussian noise instead of a homoskedastic one. 
    
    \subsection{Gaussian process surrogate with heteroskedastic nugget noise}
    In this section, the log-EDP $y(\im, \xe)$ is now supposed to follow the statistical model described by Equation \eqref{eq: stat model}
   % \begin{equation}\label{eq: stat model het}
    %    \log(\edp(\im, \xe)) = \regr(\im, \xe) + \nugget(a, \xe) \ , 
    %\end{equation}
    where $\nugget(a, \xe) \sim \mathcal{N}(0, \sigeps(a, \xe)^2)$. There are two ways of estimating $\sigeps(a, \xe)$ described in \cite{KypriotiTaflanidis2021}. The first one, called Stochastic Kriging (SK), is to consider several replications at the same value of the input parameters $(a, \xe)$ and to provide an empirical estimation of the heteroskedastic standard deviation $\sigeps(a, \xe)$. \cy{The other one} is to propose a parametric model of the noise standard deviation $\sigeps(a, \xe) = \varphi(a, \xe; \theta)$, and to calibrate the parameters vector $\theta$ using the dataset $\dataset_n = ((\im_i, \xe_i), \y(\im_i, \xe_i))_{1 \leq i \leq n}$. We decided to implement the second method with a parametric model for several reasons. \cy{As SK imposes} to be intrusive with respect to the stochastic ground motion generator in order to make several replications at precise seismic intensity $\im$, we prefer to consider to have a framework that is independent of the generator of seismic ground motions, due to the high number and diversity of stochastic generators proposed in the literature. Moreover, SK also imposes to control the design of experiments in order to be able to make replications, but in many applications, like in \cite{MarrelIoossChabridon2021}, due to budget or time constraints engineers only have access to a Monte-Carlo dataset $\dataset_n = ((\im_i, \xe_i), \y(\im_i, \xe_i))_{1 \leq i \leq n}$, this makes it impossible to perform SK to estimate the heteroskedastic noise.   
    
    The key aspect of the parametric modelization of the heteroskedastic noise is the choice of the family of parametric functions $\varphi(\im, \xe; \theta)$. A sparse representation is preferable in order to limit the dimension of the parameters vector $\theta$. Prior knowledge about earthquake engineering helps to reduce the dimension of the input parameters $(a, \xe)$. Indeed, it is common in earthquake engineering that \cy{the variability of the EDP} is mainly caused by mechanical nonlinearities \cy{and, possibly, by the variability of the seismic signals (e.g. frequency content)}, which themselves depend on the intensity of the seismic ground motion. This leads to the simplification $\varphi(a, \xe; \theta) = \varphi(a; \theta)$. Thus, $\varphi$ depends on only one variable, reducing drastically the dimension of $\theta$. The calibration of $\theta$ is performed using maximum likelihood estimation as in the homoskedastic case, $\theta$ is considered as a hyperparameter of the Gaussian process. After calibration of the hyperparameters, we can obtain the conditional distribution of the heteroskedastic Gaussian process for every $(\im, \xe)$.
    \begin{equation}\label{eq: conditional proba Y het}
        (\Ygp(\im, \xe)| \dataset_n) \sim \Norm\left(\gphetpred_n(\im, \xe), \sighetgp_n(\im, \xe)^2\right) \ ,
    \end{equation}
    we can also derive the conditional distribution of the Gaussian process $\regrgp$ on the regression function:
    \begin{equation}\label{eq: conditional proba G het}
        (\regrgp(\im, \xe)| \dataset_n) \sim \Norm\left(\gphetpred_n(\im, \xe), \siggphetlatent_n(\im, \xe)^2\right) \ ,
    \end{equation}
    where $\sighetgp_n(\im, \xe)^2 = \siggphetlatent_n(\im, \xe)^2 + \varphi(\im; \widecheck{\theta}_n)^2$, $\widecheck{\theta}_n$ is the vector of parameters of the parametrized heteroskedastic standard deviation obtained by maximum likelihood. In the same fashion as for the homoskedastic Gaussian process we can estimate the fragility curve using the conditional distribution:
    \begin{equation}\label{eq: fragepigp het}
        \fragepigphet(\im, \xe) = \Phi\left(\frac{\gphetpred_n(\im, \xe) - \log(C)}{\sighetgp_n(\im, \xe)}\right) \ ,  
    \end{equation}
    the uncertainty on the Gaussian process $(\regrgp|\dataset_n)$ can be propagated in the fragility curve in the same fashion as for the homoskedastic Gaussian process:
    \begin{align}
    \label{eq: fragepigphet uncertainty} 
            \fragepireggphet(\im, \xe) &= \Phi\bigg(\frac{\gphetreal_n(\im, \xe) - \log(C)}{\varphi(\im; \widecheck{\theta}_n)}\bigg) \ , 
    \end{align}
    where $\gphetreal_n(\im, \xe) \sim \Norm\left(\gphetpred_n(\im, \xe), \siggphetlatent_n(\im, \xe)^2\right) $.
    The distribution of $\fragepireggphet$ is empirically estimated by generating $Q$ realizations $(\gphetreal_{n, p}(\im, \xe))_{1 \leq p \leq P}$ from the distribution $\Norm\left(\gphetpred_n(\im, \xe), \siggphetlatent_n(\im, \xe)^2\right)$ in order to estimate a sample of $\fragepireggphet$:
    
    \begin{equation}\label{eq: fragepigphet uncertainty sample}
    \fragepireggphet_p(\im, \xe) = \Phi\bigg(\frac{\gphetreal_{n,p}(\im, \xe) - \log(C)}{\varphi(\im; \widecheck{\theta}_n)}\bigg) \ .
    \end{equation}

    \subsection{Uncertainty propagation on seismic fragility curves using Gaussian process surrogates}\label{sec: UQ frag}

    The Gaussian process surrogates allow us to propagate the uncertainties on $\Xe$, such that $\Xe \sim \prob_{\Xe}$, into the fragility curves by considering the random functions $\im \rightarrow \fragepi(\im, \Xe)$. We can derive from these random fragility curves several statistical quantities of interest such that the mean fragility curve:
    \begin{equation}\label{eq: mean fragility curve GP}
    \fragmean(\im) = \esp_{\Xe}[\fragepi(\im, \Xe)] \ .
    \end{equation}
    Moreover, the mechanical engineer may be interested in more conservative statistical quantities that will be useful for risk analysis. \cy{So, we define the seismic fragility quantile curve $\im \rightarrow q_{\gamma}(\im)$ of level $\gamma \in (0,1)$ as:}
    \begin{equation}\label{eq: fragility quantile}
    q_{\gamma}(\im) = \inf_{q \in \R} \big\{ \prob_{\Xe}(\fragepi(\im, \Xe) \leq q) \geq \gamma \big\} \ .
    \end{equation}
    The estimation of these quantities of interest can be carried out using a Monte-Carlo sample $(\Xe_j)_{1 \leq j \leq m}$. For the fragility quantile curve, the seismic fragility curve estimator $\fragepigp$ can be used to propose the following plug-in estimator:
    \begin{equation}\label{eq: quantile curve GP estimator}
        q^{(1)}_{\gamma}(\im) = \inf_{q \in \R} \Big\{ \frac{1}{m} \sum\limits_{j=1}^m \mathds{1}_{(\fragepigp(\im, \Xe_j) \leq q)} \geq \gamma \Big\} \ . 
    \end{equation}
    Furthermore, the posterior predictive distribution of the GP surrogates can be used to obtain the posterior distribution of the seismic fragility quantile curve using $\fragepireggp$. Using a sample of $(\fragepireggp_p)_{1 \leq p \leq P}$ of $\fragepireggp$, we can estimate a $\gamma_G$-level quantile w.r.t. the posterior distribution of the GP surrogate.
    \begin{equation}\label{eq: quantile curve GP posterior}
        q^{(2)}_{\gamma_G}(\im, \Xe) = \inf_{q \in \R} \Big\{ \frac{1}{P} \sum\limits_{p=1}^P \mathds{1}_{(\fragepireggp_p(\im, \Xe) \leq q)} \geq \gamma_G \Big\} \ . 
    \end{equation} 
    A \textit{bi-level} seismic fragility quantile curve is then proposed by taking the $\gamma_{\Xe}$-level quantile of $q^{(2)}_{\gamma_G}(\im, \Xe)$ w.r.t. the probability distribution of $\Xe$. 
        \begin{equation}\label{eq: bi-level quantile curve}
        q^{(2)}_{\gamma_G, \gamma_{\Xe}}(\im) = \inf_{q \in \R} \Big\{ \frac{1}{m} \sum\limits_{j=1}^m \mathds{1}_{(q^{(2)}_{\gamma_G}(\im, \Xe_j) \leq q)} \geq \gamma_{\Xe} \Big\} \ . 
    \end{equation} 
    The denomination bi-level meaning that it encompasses both the uncertainty on $\Xe$ and on the GP surrogate modeling. The procedure of estimation of the bi-level seismic fragility quantile curve is detailed in Algorithm \ref{alg: uncertainty propagation}. The same procedure can be applied using the heteroskedastic GP surrogate. 

    \begin{algorithm}[ht!]
        \caption{Uncertainty propagation on seismic fragility curves with Gaussian process}
        \label{alg: uncertainty propagation}
        Requirements:

        \begin{enumerate}
            \item a regular grid $(\im_t)_{1 \leq t \leq T}$
            \item a Monte-Carlo sample $(\Xe_j)_{1 \leq j \leq m}$ with the distribution of $\Xe$
            \item a learning sample $\dataset_n = ((\im_i, \xe_i), \y(\im_i, \xe_i))_{1 \leq i \leq n}$
        \end{enumerate}

        Procedure: For each $a_t$ with $1 \leq t \leq T$

        \begin{enumerate}
            \item For each $\Xe_j$ with $1 \leq j \leq m$
            \begin{enumerate}
                \item Compute with the kriging equations $\gppred_n(\im_t, \Xe_j)$, $\siggp_n(\im_t, \Xe_j)$ and $\siggplatent_n(\im_t, \Xe_j)$
                \item Compute $\fragepigp(a_t, \Xe_j)$ by Equation \eqref{eq: fragepigp estimation}
                \item For $1 \leq p \leq P$, sample $\regrgp_{n, p}(\im_t, \Xe_j) \sim \Norm(\gppred_n(\im_t, \Xe_j), \siggplatent_n(\im_t, \Xe_j)^2)$ and compute $\fragepireggp_p(\im_t, \Xe_j)$ by Equation \eqref{eq: fragepigp uncertainty sample}
            \end{enumerate}
            \item Estimate the seismic fragility quantile curve at point $a_t$ using the dataset $(\fragepigp(a_t, \Xe_j))_{1 \leq j \leq m}$ by Equation \eqref{eq: quantile curve GP estimator}
            \item Estimate the bi-level seismic fragility quantile curve with surrogate uncertainty at point $a_t$ using the dataset $(\fragepireggp_p(a_t, \Xe_j))_{\substack{1 \leq p \leq P, \\ 1 \leq j \leq m}}$ by using equations \ref{eq: quantile curve GP posterior} and \ref{eq: bi-level quantile curve}. 
        \end{enumerate}
    \end{algorithm}

    \section{Global Sensitivity Analysis of seismic fragility curves}\label{sec: GSA method}
    
    Sensitivity analysis aims at determining the input parameters of a computer model that influence the most the model response \cite{DaVeiga2021, Saltelli2007, Iooss2015}. Global Sensitivity Analysis (GSA) methods are dedicated to take into account the overall uncertainty of the input parameters. In this paper, the quantity of interest is the seismic fragility curve and thus the sensitivity index has to be defined on this quantity in order to be goal-oriented. Moreover, this is also coherent with the distinction between epistemic and aleatory uncertainties: in industrial applications, the seismic intensity measure is considered to be a penalizing input parameter which can dramatically influence the dynamical behavior of the mechanical structure studied, we thus do not need to estimate a sensitivity index of the seismic ground motion and it will not be considered as an input parameter in this part. \cy{However, the epistemic uncertainties are by definition reducible} with further data gathering or engineering studies \cite{DerKiureghian2009}. Providing accurate sensitivity indices on the input parameters tainted by epistemic uncertainties is then more justified, because they can inform the decision maker on which parameter the reduction of uncertainty will have the most impact on the quantity of interest. Thus, the global sensitivity indices will be computed only for each input parameter in $\xe$. \\  
    
    In this section, we propose two global sensitivity indices: the first ones are introduced in \cite{Iooss2019, LeGratiet2017} and coined aggregated Sobol indices, they are a natural extension of the classical Sobol indices to functional quantity of interest. The second ones are Maximum Mean Discrepancy (MMD) based Sobol indices, also coined $\beta^k$-indices \cite{Rabitz2022}. These indices are based on a reproducing kernel Hilbert Space (RKHS), thus they can handle complex types of outputs (such as functional output in our case) while being computationally tractable. 
    
    \subsection{Aggregated Sobol' indices}\label{sec agg sobol}
    Variance-based sensitivity formulation \cite{Sobol2001,Sobol1993} is a very popular way of performing GSA on computer codes, the associated sensitivity indices are coined as Sobol' indices. 
    For the case of independent inputs, we can use the ANOVA decomposition \cite{Hoeffding1948, Antoniadis1984} of a numerical model $Z = \model (X^{(1)}, ..., X^{(d)})$ where $Z$ is real-valued random variable and $(X^{(i)})_{1 \leq i \leq d}$ are $d$ real-valued random variables. The variance of $Z$ is decomposed as follows:

    \begin{equation}\label{eq: Hoeffding}
        \var(Z) = V = \sum\limits_{i=1}^d V_i + \sum\limits_{1\leq i<j \leq d} V_{ij} + \cdots + V_{1 \ldots d} =\sum_{\emptyset \varsubsetneq {\bf u}\subseteq \{1,\ldots,d\}  } V_{{\bf u}}\ ,
    \end{equation}
    where $V_i = \var\left(\esp\left[ Z | X^{(i)} \right]\right), \ V_{ij} = \var\left(\esp\left[ Z | X^{(i)}, X^{(j)} \right]\right)-  \var\left(\esp\left[ Z | X^{(i)} \right]\right) -  \var\left(\esp\left[ Z | X^{(j)} \right]\right)$,
    $V_{\bf u} =  \sum_{{\bf v} \subseteq {\bf u}} (-1)^{|{\bf u}|-|{\bf v}|}  {\rm Var}\big(\esp [ Z| (X^{(i)})_{i\in {\bf v}} ]\big) $.
    \cy{Then, the Sobol' indices are defined by}:

    \begin{equation}\label{eq: Sobol def}
        S_i = \frac{V_i}{V}, \quad S_{ij} = \frac{V_{ij}}{V}, \quad  S_{\bf u} = \frac{V_{\bf u}}{V}, \quad \ldots , \quad T_i = \sum_{ {\bf u} \subseteq \{1,\ldots,d\}, i\in {\bf u}} S_{\bf u}  \ .
    \end{equation}
    The first order Sobol' index $S_i$ measures the effect of only the input $X^{(i)}$ on the variance of the output $Z$. While the total Sobol' index $T_i$ measures the effect of $X^{(i)}$ and all the interactions between $X^{(i)}$ and the other inputs. Remark that $T_i = 1 -{V_{-i}}/{V}$ where $V_{-i} = \var\left(\esp\left[ Z | \Xe^{(-i)} \right]\right)$ and $\Xe^{(-i)} = (X^{(j)})_{j\neq i}$ the vector of all input variables except $X^{(i)}$.  

    The same kind of variance-based sensitivity indices can be defined for seismic fragility curves. It was introduced first in \cite{Iooss2019,LeGratiet2017} in the context of POD (Probability Of Detection) curves used in non destructive testing studies. Using the notation $\Xe = (\Xenum{1}, \ldots, \Xenum{d})$, we first define the following quantity:
    \begin{equation}\label{eq: agg sobol definitions}
    \begin{array}{ll}
            \frag(\im) &= \esp_{\Xe}\left[\fragepi(a, \Xe)\right] \ , \\
            & \\ 
            \fragepi_{\Xenum{i}}(a) &= \prob(\edp(\Itm, \Xe) > C | \Itm = \im, \Xenum{i}) \ , \\
            & \\ 
            \fragepi_{\Xe^{(-i)}}(a) &= \prob(\edp(\Itm, \Xe) > C | \Itm = \im, \Xe^{(-i)}) \ ,  \\
            & \\ 
            D & = \esp_{\Xe}\left[ \lVert \frag - \fragepi_{\Xe} \rVert_{L^2}^2 \right]  
            = \esp_{\Xe}\left[ \int_{a_0}^{a_1} (\frag(a) - \fragepi(a, \Xe))^2 da \right] \ ,
        \end{array}    
    \end{equation}
    where the $L^2$ norm is computed on the compact interval $\Aset = [a_0, a_1]$. Indeed, it is acceptable in terms of engineering practice to consider minimum and maximum admissible values for the seismic intensity measure. The aggregated Sobol' indices for fragility curves then write :
    \begin{align}
    \label{eq: Sobol agg 1}
            \sobol_i &= \frac{1}{D}  \int_{a_0}^{a_1} \var\left(\esp[\fragepi_{\Xe}(a)|\Xe^{(i)}]\right) da \, , \\
            \soboltot_i &=  1 - \frac{1}{D} \int_{a_0}^{a_1} \var\left(\esp[\fragepi_{\Xe}(a)|\Xe^{(-i)}]\right)  da \ ,
    \label{eq: Sobol agg 2}
    \end{align}
    where $\sobol_i$ (respectively $\soboltot_i$) is the first-order (respectively total) effect of $\Xenum{i}$ on the seismic fragility curve. These indices are coined aggregated Sobol indices because they result from the integration of the Sobol indices of the random variable $\fragepi_{\Xe}(a)$ for all admissible values of the seismic intensity measure between $a_0$ and $a_1$. Moreover, as the classical Sobol indices, they follow an ANOVA decomposition, allowing for a clear definition of the relative influence of each subset of input parameters into the seismic fragility curve uncertainty. Pick-freeze estimators \cite{DaVeiga2021} of the aggregated Sobol indices are used in order to avoid a double Monte-Carlo loop. Moreover, the Gaussian process surrogate model is used to replace the different fragility curves defined in Equation \eqref{eq: agg sobol definitions} by their estimators $\fragepigp$. Let $\Xepf$ be an independent copy of $\Xe$. We define $\Xepf_i = (\Xenumpf{1},\ldots, \Xenum{i}, \ldots, \Xenumpf{d})$ and $\Xepf_{-i} = (\Xenum{1},\ldots, \Xenumpf{i}, \ldots, \Xenum{d})$. 
    The pick-freeze principle relies on the following result, for all $a \in [a_0, a_1]$:
    \begin{equation}\label{eq: pick-freeze principle}
            \var\left(\esp[\fragepi_{\Xe}(a)|\Xe^{(i)}]\right) = \cov(\fragepi_{\Xe}(a), \fragepi_{\Xepf_i}(a)) \ .  \\
    \end{equation}
    By plugging Equation \eqref{eq: pick-freeze principle} into Equations \eqref{eq: Sobol agg 1}-\ref{eq: Sobol agg 2}, it is possible to define a pick-freeze estimator of the aggregated Sobol indices. We draw a Monte-Carlo sample of size $m$ of $(\Xe,\Xepf_i,\Xepf_{-i})$ that we denote $(\Xe_j,  \Xepf_{i, j}, \Xepf_{-i, j})_{1 \leq j \leq m}$. The aggregated Sobol indices estimators then write:
    \begin{align}
    \label{eq: Sobol estimators 1}
            \sobolestim_{i, m, n} &= \frac{\displaystyle \sum\limits_{t=1}^T \left\langle \fragepigp(a_t, \Xe)\fragepigp(a_t, \Xepf_{i}) \right\rangle_m - \left\langle  \fragepigp(a_t, \Xe)\right\rangle_m \left\langle  \fragepigp(a_t, \Xepf_{i})\right\rangle_m}{\displaystyle \sum\limits_{t=1}^T \left\langle \fragepigp(a_t, \Xe)^2 \right\rangle_m - \left\langle \fragepigp(a_t, \Xe) \right\rangle_m^2} \, , \\
            \soboltotestim_{i, m, n} &= 1 -  \frac{\displaystyle \sum\limits_{t=1}^T \left\langle \fragepigp(a_t, \Xe)\fragepigp(a_t, \Xepf_{-i}) \right\rangle_m - \left\langle \fragepigp(a_t, \Xe)\right\rangle_m \left\langle \fragepigp(a_t, \Xepf_{-i})\right\rangle_m}{\displaystyle \sum\limits_{t=1}^T \left\langle \fragepigp(a_t, \Xe)^2\right\rangle_m - \left\langle\fragepigp(a_t, \Xe)\right\rangle_m^2} \, ,
    \label{eq: Sobol estimators 2}
    \end{align}
where we denote for any function $f$:
\begin{equation}
    \left< f(\Xe,\Xepf_i,\Xepf_{-i}) \right>_m = 
    \frac{1}{m}\sum_{j=1}^m  f(\Xe_j,\Xepf_{i,j},\Xepf_{-i,j}) \, .
\end{equation}
    A regular grid $(a_t)_{1 \leq t \leq T}$ is used to approximate the integral on the seismic intensity measure on the set $[a_0, a_1]$.
    
    Given that the Gaussian process surrogate provides a predictor and its associated uncertainty, we can propagate it into the aggregated Sobol' indices estimators by replacing the fragility curve estimator $\fragepigp$ by $P$ draws $\fragepireggp_p$ using the probability distribution of $(G|\dataset_n)$. The aggregated Sobol indices then write: 
    
    \begin{align}
    \label{eq: Bayesian Sobol estimators 1}
            \sobolestimgp_{i, m, n, p} &= \frac{\displaystyle \sum\limits_{t=1}^T \left\langle \fragepireggp_p(a_t, \Xe)\fragepireggp_p(a_t, \Xepf_{i}) \right\rangle_m - \left\langle \fragepireggp_p(a_t, \Xe)\right\rangle_m \left\langle \fragepireggp_p(a_t, \Xepf_{i})\right\rangle_m}{\displaystyle \sum\limits_{t=1}^T \left\langle  \fragepireggp_p(a_t, \Xe)^2 \right\rangle_m - \left\langle \fragepireggp_p(a_t, \Xe)\right\rangle_m^2} \, , \\
            \soboltotestimgp_{i, m, n, p} &= 1 -  \frac{\displaystyle \sum\limits_{t=1}^T \left\langle \fragepireggp_p(a_t, \Xe)\fragepireggp_p(a_t, \Xepf_{-i}) \right\rangle_m - \left\langle \fragepireggp_p(a_t, \Xe)\right\rangle_m \left\langle \fragepireggp_p(a_t, \Xepf_{-i})\right\rangle_m}{\displaystyle \sum\limits_{t=1}^T \left\langle \fragepireggp_p(a_t, \Xe)^2 \right\rangle_m - \left\langle \fragepireggp_p(a_t, \Xe)\right\rangle_m^2} \, . 
        \label{eq: Bayesian Sobol estimators 2}
    \end{align}

    The aggregated Sobol' indices estimators defined in Equations \eqref{eq: Sobol estimators 1}-\eqref{eq: Sobol estimators 2} use the GP predictor of the fragility curve $\fragepigp$ to quantify the impact of each input parameter on the overall fragility curve. The uncertainty on the regression function is obtained with the probability distribution of $(\regrgp|\dataset_n)$ or $(\gphetreal|\dataset_n)$ and is propagated into the aggregated Sobol' indices estimators in Equations \eqref{eq: Bayesian Sobol estimators 1}-\ref{eq: Bayesian Sobol estimators 2}. Moreover, in order to take into account the uncertainty of the Monte-Carlo estimation of the Sobol indices, we draw, for $b=1,\ldots,B$, the random variables  $(u_b(j))_{1 \leq j \leq m}$ with equiprobability and with replacement in $\{1,\ldots,m\}$ and replace the pick-freeze Monte-Carlo sampling dataset $(\Xe_j, \Xepf_{i, j},\Xepf_{-i, j})_{1 \leq j \leq m}$ by $(\Xe_{u_b(j)}, \Xepf_{i, u_b(j)}, \Xepf_{-i, u_b(j)})_{1 \leq j \leq m}$. We thus obtain a sample of size $P \times B$ of aggregated Sobol indices $(\sobolestimgp_{i, m, n, p, b})_{\substack{1 \leq p \leq P, \\ 1 \leq b \leq B}}$. This sample allows us to quantify the uncertainty of $\sobol_i$ coming from the kriging metamodel uncertainty and the pick-freeze Monte-Carlo uncertainty. The same procedure can be made with total Sobol indices $\soboltot_i$. The estimation of the metamodel and Monte-Carlo uncertainty on $\sobol_i$ is presented in Algorithm \ref{alg: posterior distrib agg sobol}. Note that for pick-freeze estimators we have to sample Gaussian vectors of size $2mT$ which is in our application close to $10^6$. The classical sampling method for sampling Gaussian vectors use a Cholesky decomposition of the covariance matrix and has a cubic complexity with respect to the Gaussian vector size. Here we use sampling by kriging conditioning and Nyström procedure as described in \cite{LeGratiet2014} to make the computations tractable. In the same manner as in \cite{LeGratiet2014}, we can also estimate the part of variance of $\sobol_i$ coming from the Monte-Carlo approximation and the part related to the kriging metamodel uncertainty. The part of variance related to the metamodeling writes:
    \begin{equation}\label{eq: kriging sobol variance}
    \widehat{\sigma}_{\regrgp_n}^2(\sobolestimgp_{i, m, n}) = \frac{1}{B} \sum\limits_{b=1}^B \frac{1}{P - 1}\sum\limits_{p=1}^P \big(\sobolestimgp_{i, m, n, p, b} - 
    \left<\sobolestimgp_{i, m, n, b} \right>_P
%    \sobolestimgpbar_{i, m, n, b}
    \big)^2 \, , 
    \end{equation}
    where $ %\sobolestimgpbar_{i, m, n, b} 
    \left<\sobolestimgp_{i, m, n, b} \right>_P=  \frac{1}{P}  \sum\limits_{p=1}^P \sobolestimgp_{i, m, n, p, b}$. 
    Furthermore, it is also possible to evaluate the part of the variance due to Monte-Carlo approximation of the aggregated Sobol indices:
    \begin{equation}\label{eq: MC sobol variance}
    \widehat{\sigma}_{{\rm MC}_m}^2(\sobolestimgp_{i, m, n}) = \frac{1}{P} \sum\limits_{p=1}^P \frac{1}{B - 1}\sum\limits_{b=1}^B \big(\sobolestimgp_{i, m, n, p, b} - 
    \left< \sobolestimgp_{i, m, n, p}\right>_B
    \big)^2 \, ,
    \end{equation}
    where $%\sobolestimgpbarbar_{i, m, n, k} 
        \left< \sobolestimgp_{i, m, n, p}\right>_B
        = \frac{1}{B} \sum\limits_{b=1}^B \sobolestimgp_{i, m, n, p, b}$. Following \cite{LeGratiet2014}, we can use these two variances as a rationale for choosing the number of Monte Carlo samples $m$ and the number of mechanical simulations of the structure $n$. Indeed, when $\widehat{\sigma}_{{\rm MC}_m}^2(\sobolestimgp_{i, m, n}) \approx \widehat{\sigma}_{\regrgp_n}^2(\sobolestimgp_{i, m, n})$ the Monte Carlo and the kriging metamodel errors have the same contributions into the estimation error of the aggregated Sobol indices. Remark that these variances are defined for each input parameter and each order of the aggregated Sobol indices. A compromise has to be made for choosing which order and input parameter the engineer must consider.  

    \begin{algorithm}
    \caption{Estimation of the metamodel and Monte-Carlo uncertainty on $\sobol_i$}
    \label{alg: posterior distrib agg sobol}
         Same requirements as Algorithm \ref{alg: uncertainty propagation}, with additionally:
        \begin{enumerate}
%            \item the posterior distribution $(\regrgp|\dataset_n)$,
            \item a Monte-Carlo sample $(\Xe_j, \Xepf_{i, j})_{1 \leq j \leq m}$ with the distribution of $(\Xe, \Xepf_{i})$
            \item the number $P$ of realizations of the GP posterior distribution
            \item the number $B$ of bootstrap samples
        \end{enumerate}

        Procedure: For $1 \leq p \leq P$
    \begin{enumerate}
        \item Sample $\regrgp_{n, p}(\mathbf{D})$ with the posterior distribution $(\regrgp(\mathbf{D})|\dataset_n)$, where $\mathbf{D} = (a_t, \Xe_j)_{1 \leq t \leq T, 1 \leq j \leq m} \cup (a_t, \Xepf_{i, j})_{1 \leq t \leq T, 1 \leq j \leq m}$
        \item For $1 \leq b \leq B$
        \begin{enumerate}
                \item Sample with replacement in $\{1,\ldots,m\}$ the bootstrap indices $(u_b(j))_{1 \leq j \leq m}$ and then define the bootstrap sample $\mathbf{D}^b = (a_t, \Xe_{u_b(j)})_{1 \leq t \leq T, 1 \leq j \leq m} \cup (a_t, \Xepf_{i, u_b(j)})_{1 \leq t \leq T, 1 \leq j \leq m}$
            \item Compute $\sobolestimgp_{i, m, n, p,b}$ by Equations \eqref{eq: fragepigp uncertainty sample} and \eqref{eq: Bayesian Sobol estimators 1} and by using $\regrgp_{n, p}(\mathbf{D}^b)$
        \end{enumerate}
    \end{enumerate}
\end{algorithm}
    
    \subsection{$\beta^k$ indices}
    
    Kernel-based methods in machine learning and statistics gains in popularity due to their ability to simplify difficult nonlinear problems into linear problems by embedding the data points into a Reproducing Kernel Hilbert Space (RKHS) \cite{Scholkopf2002}. The main applications involve independence testing \cite{Gretton2007, Fukumizu2007} and dimension reduction \cite{Scholkopf1998, Fukumizu2004, Fukumizu2009}. A first use of kernel methods for GSA purposes was proposed in \cite{DaVeiga2015} where the Hilbert Schmidt Independence Criterion (HSIC) is used to propose global sensitivity indices. $\beta^k$ indices \cite{Rabitz2022} make also use of the RKHS and are global sensitivity indices based on the Maximum Mean Discrepancy (MMD) and defined with the rationale of \cite{Borgonovo2016}. $\beta^k$ indices have also the interesting property of being Sobol indices on the kernel embedding of the output variable as shown in \cite{DaVeiga2021}. This is appealing for our application as we can use the same framework of estimation as for the aggregated Sobol indices using a pick-freeze scheme and propagate the kriging prediction uncertainty into the $\beta^k$ indices estimates. In order to define the $\beta^k$ indices, we have to define the MMD given the kernel function $(x,y) \mapsto k(x, y)$ \cite{Gretton2012}.
    
    \begin{lemma}
    Let $\inputspace$ be a separable topological nonempty set and $\probT$ be the set of all probability measures on $\inputspace$. Let $(u, v) \mapsto k(u, v)$ be a continuous positive-definite kernel. Let $\mathbb{P}, \mathbb{Q} \in \probT$. Suppose $U, U' \sim \mathbb{P}$ and $V, V' \sim \mathbb{Q}$, where $U, \ U', \ V, \ V'$ are mutually independent, such that $\esp[\sqrt{k(U, U')}] < +\infty$ and $\esp[\sqrt{k(V, V')}] < +\infty$. The Maximum Mean Discrepancy (MMD) between $\mathbb{P}$ and $\mathbb{Q}$ can be expressed as follows:
    \begin{equation}
    \label{eq: MMD def}
        \mmd(\mathbb{P}, \mathbb{Q})^2 = \esp[k(U, U')] + \esp[k(V, V')] - 2\esp[k(U, V)] \, .
    \end{equation}
    \end{lemma}

    The MMD allows us to define a distance between probability measures. According to \cite{DaVeiga2015, Borgonovo2016}, it is possible to define a global sensitivity index using a distance between probability measures. Given the same numerical model as in Section \ref{sec agg sobol} $Z = \model(X^{(1)},\ldots,X^{(d)})$, the $\beta^k$ index for the input variable $X^{(i)}$ is defined by:
    \begin{equation}\label{eq: betak def}
        \beta^k_i = \frac{\esp_{X^{(i)}}[\mmd(\mathbb{P}_{Z}, \mathbb{P}_{Z|X^{(i)}})^2]}{\esp_{\Xe}[\mmd(\mathbb{P}_{Z}, \mathbb{P}_{Z|\Xe})^2]} \ .
    \end{equation}
    
    As shown in \cite{DaVeiga2021}, these indices follow an ANOVA decomposition, the relative influence of each group of input parameters can be assessed. We can then define the total order $\beta^k$ index for the variable $X^{(i)}$ as follows:
    \begin{equation}\label{eq: betak tot def}
        \beta^k_{-i} = 1 - \frac{\esp_{\Xe}[\mmd(\mathbb{P}_{Z}, \mathbb{P}_{Z|\Xe^{(-i)}})^2]}{\esp_{\Xe}[\mmd(\mathbb{P}_{Z}, \mathbb{P}_{Z|\Xe})^2]} \ .
    \end{equation}
    
    In the same spirit of Section \ref{sec agg sobol}, it is possible to estimate the $\beta_k$-indices using a pick-freeze estimation framework as shown in \cite{DaVeiga2021}. Using the same notations as in Section \ref{sec agg sobol}, we can rewrite the first order $\beta^k$ index as: 
    \begin{equation}\label{eq: betak pick-freeze 1}
        \beta^k_i = \frac{\esp[k(\model(\Xe), \model(\Xepf_i))] - \esp[k(\model(\Xe), \model(\Xepf))]}{\esp[k(\model(\Xe), \model(\Xe))] - \esp[k(\model(\Xe), \model(\Xepf))]} \ ,
    \end{equation}
    the total order $\beta^k$ index can be also expressed as:
   \begin{equation}\label{eq: betak pick-freeze 2}
        \beta^k_{-i} = 1 - \frac{\esp[k(\model(\Xe), \model(\Xepf_{-i}))] - \esp[k(\model(\Xe), \model(\Xepf))]}{\esp[k(\model(\Xe), \model(\Xe))] - \esp[k(\model(\Xe), \model(\Xepf))]} \ .
    \end{equation}
   
    After defining the $\beta^k$ indices, we have to adapt to the case where the output variable of interest is no longer a scalar variable but a functional variable. In order to define $\beta^k$ indices on the seismic fragility curves, define $\mathcal{F} = L^2([a_0, a_1])$. We thus have to define a positive definite kernel on $\mathcal{F} \times \mathcal{F}$, $(\Psi_1,\Psi_2) \rightarrow \kernelfrag(\Psi_1,\Psi_2)$ for $\Psi_1,\Psi_2 \in \mathcal{F}$. According to \cite{Ferraty2006}, let $\Delta(.,.)$ be a semi-metric defined on the functional space $\mathcal{F} \times \mathcal{F}$, a kernel associated to $\mathcal{F}$ can be defined as $\kernelfrag(\Psi_1, \Psi_2) = k(\Delta(\Psi_1, \Psi_2))$ where $k$ is acting on $\mathbb{R}$. For the sake of notations simplicity, the kernel acting on the functional space $\mathcal{F}$ will be denoted by $k$. For our application, we will choose the so called Gaussian kernel with squared $L^2$ norm:
    \begin{equation}\label{eq: kernel frag gaussian}
        k(\Psi_1, \Psi_2) = \exp\left( - \frac{\lVert \Psi_1 - \Psi_2 \rVert_{L^2}^2}{2\ell^2} \right) \ ,
    \end{equation}
    where $\ell$ is a hyperparameter of the kernel that will be calibrated with the available data. The pick-freeze method combined with the Gaussian process surrogates allows us to define the following $\beta^k$ indices estimators:
    \begin{align}
    \label{eq: betak pick-freeze estimator 1}
            \betakestim_{i, m} &= \frac{\displaystyle \left\langle k\left(\fragepigp(., \Xe), \fragepigp(., \Xepf_{i})\right) - k\left(\fragepigp(., \Xe), \fragepigp(., \Xepf)\right) \right\rangle_m}{\displaystyle \left\langle k\left(\fragepigp(., \Xe), \fragepigp(., \Xe)\right) - k\left(\fragepigp(., \Xe), \fragepigp(., \Xepf)\right) \right\rangle_m} \, , \\
            \betakestim_{-i, m} &= 1 -  \frac{\displaystyle \left\langle k\left(\fragepigp(., \Xe), \fragepigp(., \Xepf_{-i})\right) - k\left(\fragepigp(., \Xe), \fragepigp(., \Xepf)\right) \right\rangle_m}{\displaystyle \left\langle k\left(\fragepigp(., \Xe), \fragepigp(., \Xe)\right) - k\left(\fragepigp(., \Xe), \fragepigp(., \Xepf)\right) \right\rangle_m} \ .
    \label{eq: betak pick-freeze estimator 2}
    \end{align}
    
    In the same fashion as for the aggregated Sobol indices it is possible to propagate the uncertainty of the posterior distribution $(\regrgp|\dataset_n)$ of the Gaussian process using $P$ realizations:
    \begin{align}
    \label{eq: Bayesian betak pick-freeze estimator 1}
        \betakestimgp_{i, m, p} = \frac{\displaystyle \left\langle k\left(\fragepireggp_p(., \Xe), \fragepireggp_p(., \Xepf_{i})\right) - k\left(\fragepireggp_p(., \Xe), \fragepireggp_p(., \Xepf)\right) \right\rangle_m}{\displaystyle \left\langle k\left(\fragepireggp_p(., \Xe), \fragepireggp_p(., \Xe)\right) - k\left(\fragepireggp_p(., \Xe), \fragepireggp_p(., \Xepf)\right) \right\rangle_m} \, , \\
        \betakestimgp_{-i, m, p} = 1 -  \frac{\displaystyle \left\langle k\left(\fragepireggp_p(., \Xe), \fragepireggp_p(., \Xepf_{-i})\right) - k\left(\fragepireggp_p(., \Xe), \fragepireggp_p(., \Xepf)\right) \right\rangle_m}{\displaystyle \left\langle k\left(\fragepireggp_p(., \Xe), \fragepireggp_p(., \Xe)\right) - k\left(\fragepireggp_p(., \Xe), \fragepireggp_p(., \Xepf)\right) \right\rangle_m}\ .
           \label{eq: Bayesian betak pick-freeze estimator 2}
\end{align}

Remark that - similarly to the aggregated Sobol indices - we can estimate the share of variance of the $\beta^k$ indices estimators due to the Monte-Carlo pick-freeze estimation method and due to the Gaussian process surrogate model uncertainty using Equations \eqref{eq: MC sobol variance} and \ref{eq: kriging sobol variance}. 
    
    \section{Application to a safety water pipe of a French PWR}\label{sec: numerical appli}

    \subsection{Presentation of the use case}
    
\begin{figure}[!ht]
		\centering		
		\subfloat[\label{fig:ASG_MU}]{\includegraphics[width=6cm]{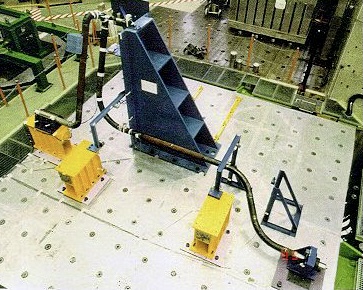}}
		\hspace{0.5cm}
		\subfloat[\label{fig:ASG_FEM}]{\includegraphics[trim= 1cm  3.8cm 12cm 1.5cm, clip,width=5cm]{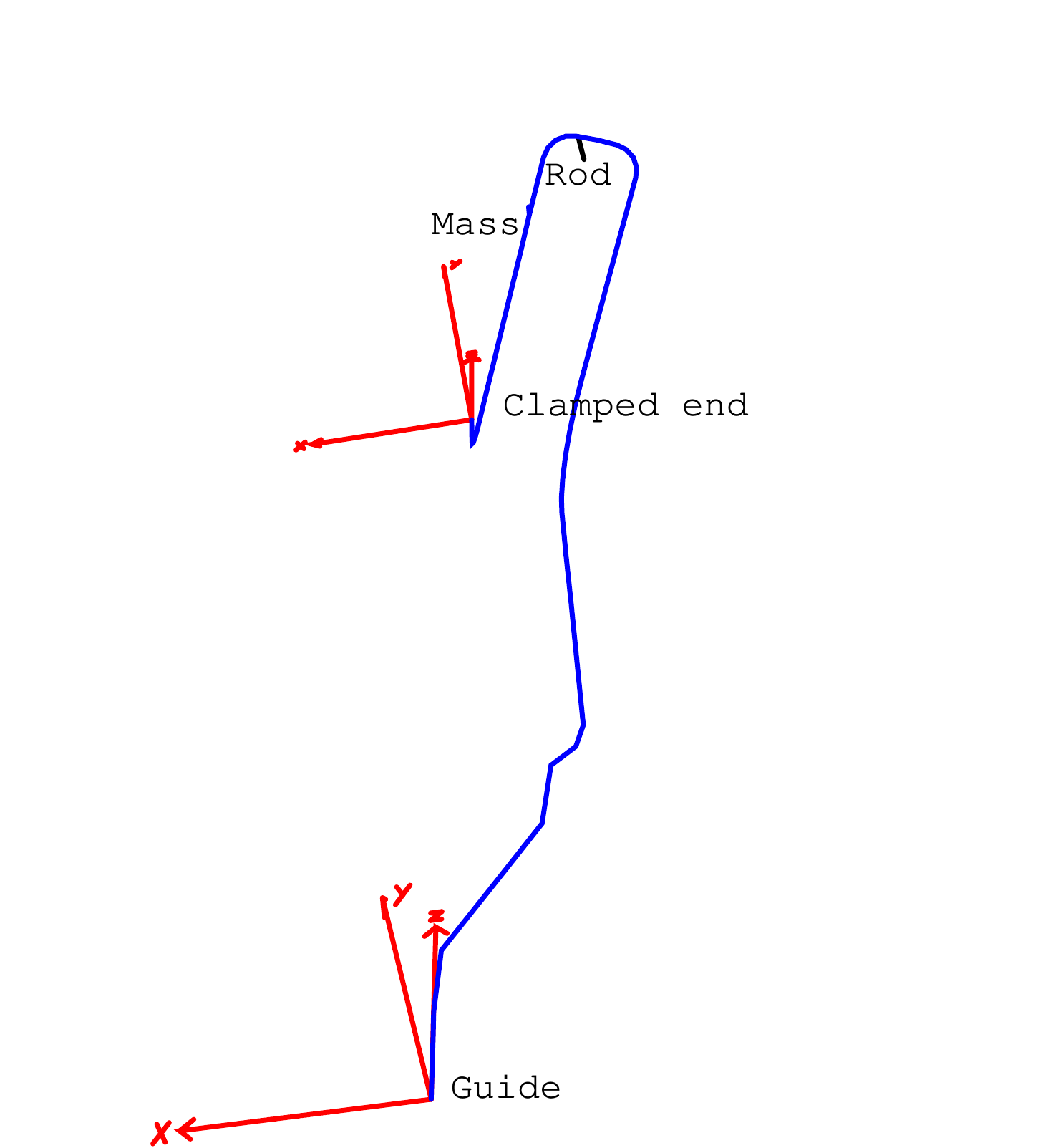}}
		\caption{(a) Overview of the ASG mock-up on the CEA's shaking table and (b) ASG FE model.}
		\label{fig:ASG}
	\end{figure}

    \cy{Regulatory seismic risk prevention work for the nuclear power plants includes the study of piping systems. Thus, this use case is related to a numerical model of a part of a piping system which was validated after an experimental campaign on a mock-up based on seismic tests on the Azalee shaking table of the EMSI laboratory of CEA Saclay}. The main results of this experimental program, called ASG program, are detailed in \cite{Touboul1999}. The Finite Element (FE) model, based on beam elements, is implemented with the homemade FE code CAST3M \cite{CAST3M}. In Figure \ref{fig:ASG_MU} a view of the mock-up mounted on the shaking table is shown. The FE model is depicted in Figure \ref{fig:ASG_FEM}.

    The output variable of interest is the maximum of the out-of-plane rotation of a specific elbow of the piping system. \cy{This is the EDP of this problem, as recommended in \cite{TOUBOUL2006}. The sources of epistemic uncertainties are the mechanical parameters of the numerical model and the boundary conditions, in order to take into account that the mock-up is in practice part of a much larger piping system. In our setting, the mass of the mock-up is considered as perfectly known}. The 10 \cy{uncertain} parameters are detailed in Table \ref{table:variablesASG}. \cy{All the associated random variables} follow uniform probability distributions with prescribed means (the numerical values are in Table \ref{table:variablesASG}) and with coefficients of variation of $15\%$. All inputs are considered mutually independent. The mean value of each parameter is calibrated in the following manner: The mock-up is part of a bigger piping system with a known first eigenmode obtained through numerical simulations, so we choose the mean value for the boundary condition's parameters so that the first eigenmode of the mock-up matches the first eigenmode of the mock-up when coupled to the entire piping system. Therefore, computational experiments based on simulations with calibrated mock-up boundary conditions are more representative of the mock-up in its real environment.

    \begin{table}[!ht]
        \caption{Epistemic variables definition for the ASG use case.}
        \label{table:variablesASG}
        \begin{center}
            \resizebox{\columnwidth}{!}{
            \begin{tabular}{|c|c|c|} 
            \hline
            \textit{Variable number} & \textit{Variable} & \textit{Mean}\\ 
            \hline 
             $1$ & E, Young modulus & $1.9236 \ 10^{11}$ Pa \\ 
             $2$ & Sy, Elasticity limit & $300$ MPa \\
             $3$ & H, Hardening module & $4.27 \ 10^{8}$  \\
             $4$ & b, Modal damping ratio & $1\%$ \\
             $5$ & RPY151, Rotation stiffness for the P151 guide in Y direction & $1.1 \ 10^{5}$ Nm/rad \\
             $6$ & RPX29, Rotation stiffness for the P29 clamped end in X direction & $1.1 \ 10^{5}$ Nm/rad \\ 
             $7$ & RPY29, Rotation stiffness for the P29 clamped end in Y direction & $3.3 \ 10^{5}$ Nm/rad \\ 
             $8$ & TPX29, Translation stiffness for the P29 clamped end in X direction & $1.0 \ 10^{6}$ N/m \\
             $9$ & TPY29, Translation stiffness for the P29 clamped end in Y direction & $2.0 \ 10^{5}$ N/m \\
             $10$ & TPZ29, Translation stiffness for the P29 clamped end in Z direction & $1.0 \ 10^{6}$ N/m \\
             \hline
            \end{tabular}}
        \end{center}
    \end{table}

    %The input parameters are the same for both models. 
    Due to the limited number of records of real seismic ground motions acceleration signals, it is common to generate artificial seismic signals using a stochastic generator fitted on real accelerogram records. \cy{We use the stochastic generator defined in \cite{Rezaeian10} whose calibration is described in \cite{Sainct2020}. Finally, as the piping system is in practice located in a building, the synthetic signals are filtered by a deterministic fictitious linear single-mode building at $5$ Hz and damped at $2\%$.}
    %The parameters are drawn according to the probability distributions described in Table \ref{table:variablesASG}. 

    The computer model of the ASG mock-up is composed of a linear FE model when the maximal stress in the mock-up pipe elbow is less than the elasticity limit Sy and a nonlinear FE model when the maximal stress is greater. A run of the linear FE model has a computation of a dozen of seconds \cy{- the numerical resolution is based on a modal based projection -} whereas a run of the nonlinear FE model has a computation time of approximately ten minutes. 
    
    \subsection{Dimension reduction of the input space and choice of the heteroskedastic noise parametric model}
    
    \cy{In this section, we present a data selection step to reduce the dimension of the input space of the mechanical computer model and the choice of the model of the variance that is retained for the heteroskedastic GP.}
    
    The dimension reduction step of the input space was performed with a HSIC based statistical hypothesis test using the ICSCREAM methodology developed in \cite{MarrelIoossChabridon2021}: a Gaussian kernel was used for each input variable and for the output variable (i.e. the log rotation of the pipe elbow). \cy{2000 mechanical simulations using the less expensive linear FE model were carried out for the mechanical input variables screening and 6 variables were selected} (the variables number $1,2,3,8,9,10$ in Table \ref{table:variablesASG}).
    
    For the parametric form of the standard deviation for the heteroskedastic Gaussian process, we consider the following ramp function:
    \begin{equation}\label{eq: het parametric nugget}
        \varphi(\im; \theta) = \max(\theta_0 + \theta_1 a, \theta_2) \ , 
    \end{equation}
    where $\theta = (\theta_0, \theta_1, \theta_2)$. This parametric model for the heteroskedastic standard deviation is motivated by the model proposed in \cite{KypriotiTaflanidis2021}. It has the advantage to depend only on one variable and the small dimension of $\theta$ allows for its calibration with a reasonable sized dataset ($n < 1000$).
    
    For the homoskedastic Gaussian process model, the hyperparameters are estimated using the maximum a posteriori estimator proposed in \cite{Gu2018} using a so-called jointly robust prior, which has the useful property to avoid hyperparameters values raising ill-conditioned correlation matrices. On the other hand, the heteroskedastic Gaussian process \cy{are} estimated using maximum likelihood. A Monte-Carlo sample of size $n=500$ from the probability distribution of the mechanical parameters $\Xe$ is drawn, as well as $500$ realizations of our stochastic ground motion generator model. The \cy{$n = 500$ mechanical simulations are} then carried out using CAST3M. The performance of the homoskedastic and heteroskedastic GP models is then assessed in the following section.

    \subsection{Performance evaluation of the Gaussian process surrogates}
    
   \cy{This section is devoted to the qualitative evaluation of the predictive properties of the two surrogate models.}
    
\begin{figure}[!ht]
        \centering
        \begin{subfigure}[b]{0.48\textwidth}
         \centering
         \includegraphics[width=\textwidth]{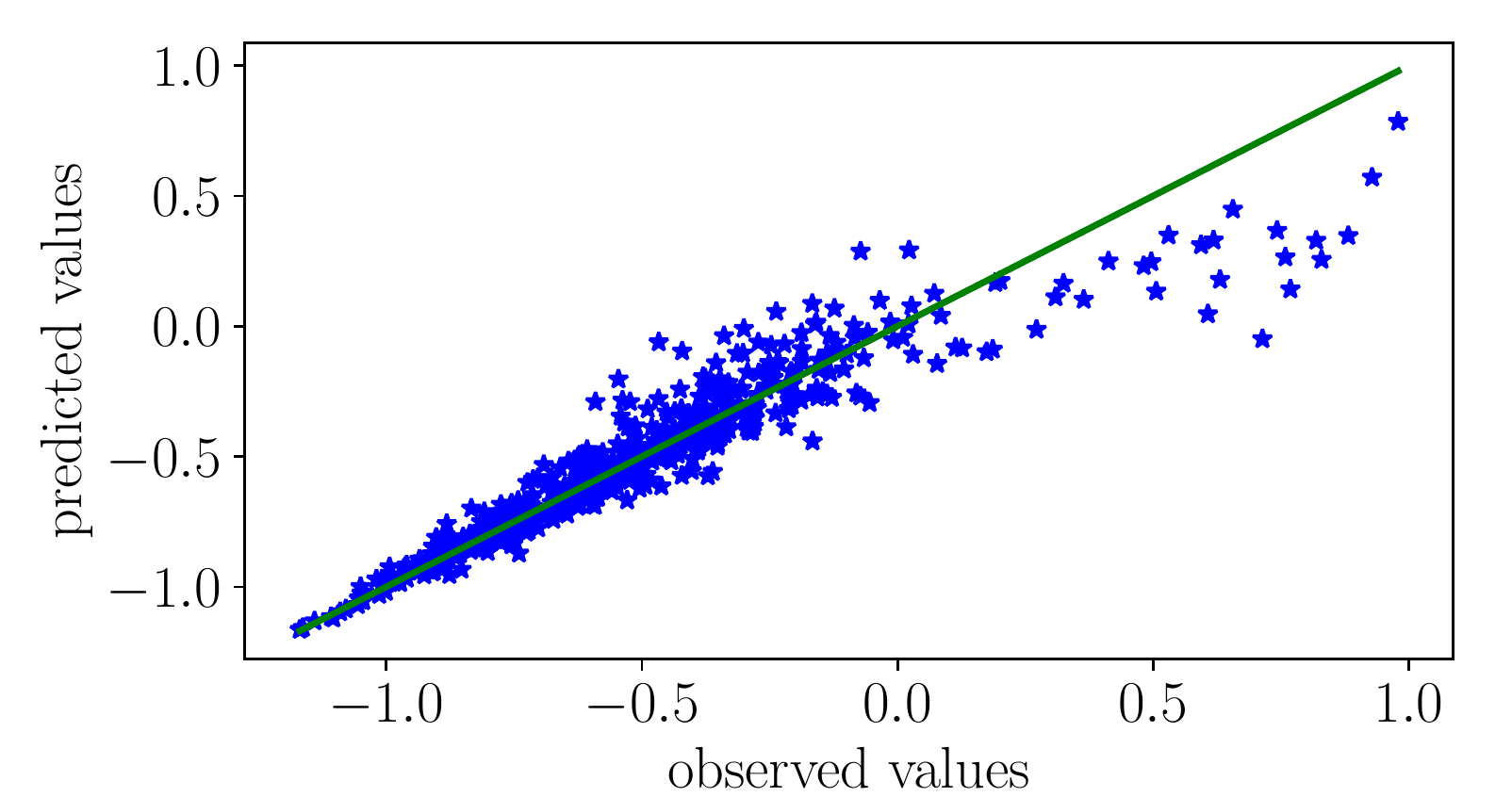}
         \caption{Heteroskedastic Gaussian process} 
     \end{subfigure}
     \hfill
     \begin{subfigure}[b]{0.48\textwidth}
         \centering
         \includegraphics[width=\textwidth]{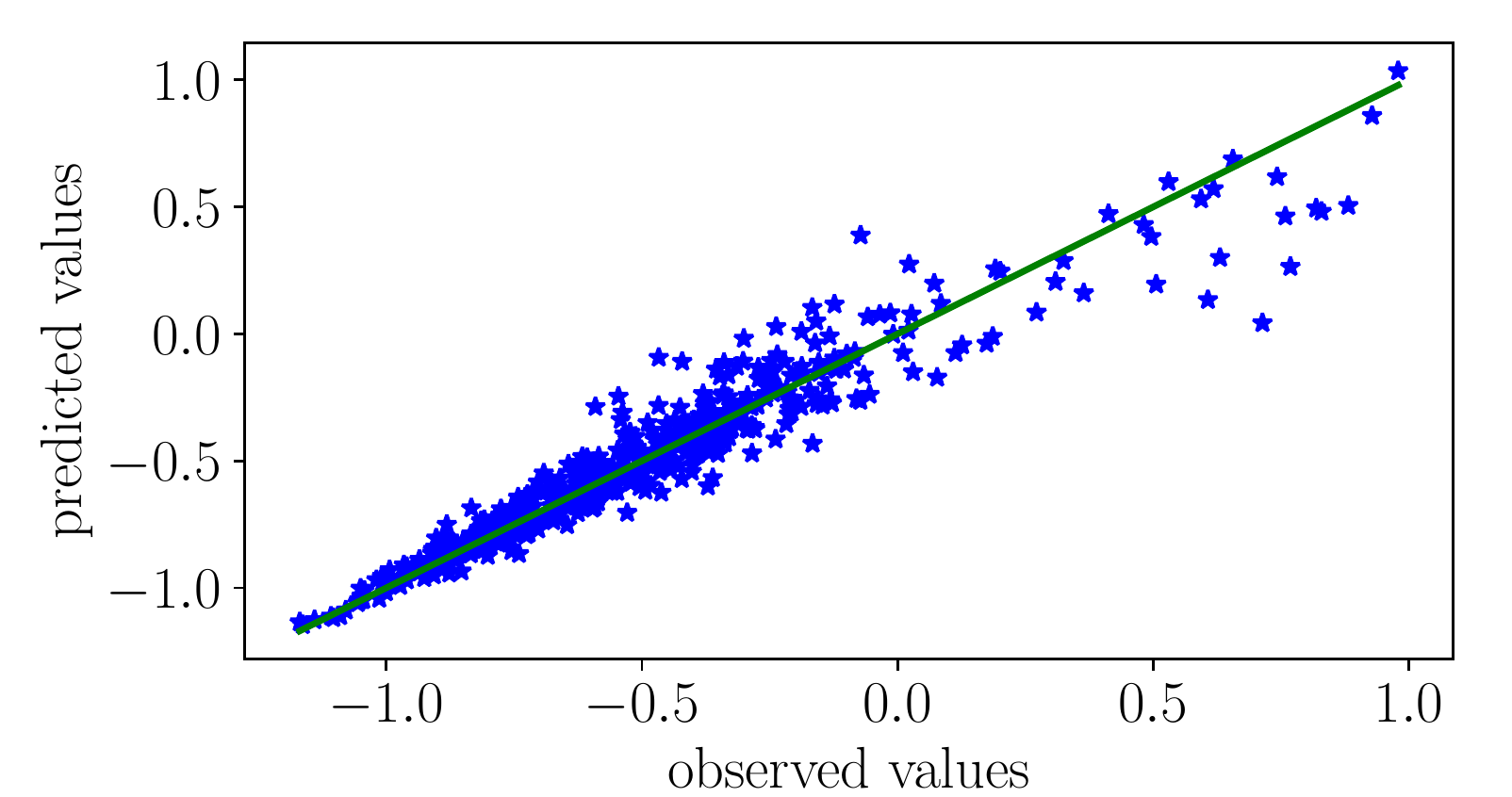}
         \caption{Homoskedastic Gaussian process}  
     \end{subfigure}
     \caption{\label{fig: predicted observed} Predicted values versus observed values for the heteroskedastic and homoskedastic Gaussian process surrogate with a dataset size $n=500$.}
\end{figure}

    Figure \ref{fig: predicted observed} shows the predicted versus observed values of the log-EDP $\y(\im, \xe)$ using a learning dataset of $n=500$ observations. The green solid line corresponds to the identity, the closer the data are from this line the better the prediction quality of the surrogate is. We can notice that the heteroskedastic Gaussian process underestimates the high values of the log-EDP, the homoskedastic surrogate have also this behavior but the data are closer to the identity line for high values of the log-EDP. However, this concerns the log-EDP values greater than the 90 \%-level quantile. Hence, it is not sufficient to determine whether the homoskedastic or the heteroskedastic Gaussian process has the best performance in terms \cy{of} prediction.    
    
    In order to study more quantitatively the predictive properties of the two surrogates, we use the prediction power $Q^2$ defined as:
    
    \begin{equation}\label{eq: q2}
        Q^2 = 1 - \frac{\sum_{i=1}^{n_t} (y(\im_i^{t}, \xe_i^{t}) - \gppred_n(\im_i^{t}, \xe_i^{t}))^2}{\sum_{i=1}^{n_t} (y(\im_i^{t}, \xe_i^{t}) - \bar{y})^2} \ ,
    \end{equation}
    where $(\im_i^{t},\xe_i^{t}, y(\im_i^{t}, \xe_i^{t}))_{1 \leq i \leq n_t}$ is a test dataset, and $\bar{y} = \frac{1}{n_t} \sum_{i=1}^{n_t} y(\im_i^{t},\xe_i^{t})$. In practice the computational cost of mechanical \cy{models} limits the sample size, thus the prediction power $Q^2$ is computed using leave-one-out cross validation technique \cite{Dubrule1983}: The hyperparameters of the GP \cy{surrogates} are estimated only once on the training sample to alleviate the computational burden of hyperparameter tuning in the cross-validation procedure. Table \ref{tab: q2 values loo} \cy{gathers} the $Q^2$ numerical values for the homoskedastic and heteroskedastic GP and learning sample size between 100 and 500. The $Q^2$ values of the heteroskedastic and homoskedastic GP surrogates being very close to each other, we can conclude that the two surrogates raise the same predictive performance.     

    \begin{table}[ht]
    \centering
    \begin{tabular}{c|c|c|c|c|c}
        Learning sample size & 100 & 200 & 300 & 400 & 500 \\ \hline
        Homoskedastic & $0.844$ & $0.860$ & $0.853$ & $0.870$ & $0.867$ \\ 
        Heteroskedastic & $0.842$ & $0.860$ & $0.849$ & $0.872$ & $0.875$ 
    \end{tabular}
    \caption{$Q^2$ numerical values estimated by leave-one-out on the training sample for various learning sample size and the two GP surrogates.}
    \label{tab: q2 values loo}
\end{table}

\begin{figure}[!ht]
     \centering
     \begin{subfigure}[b]{0.48\textwidth}
         \centering
         \includegraphics[width=\textwidth]{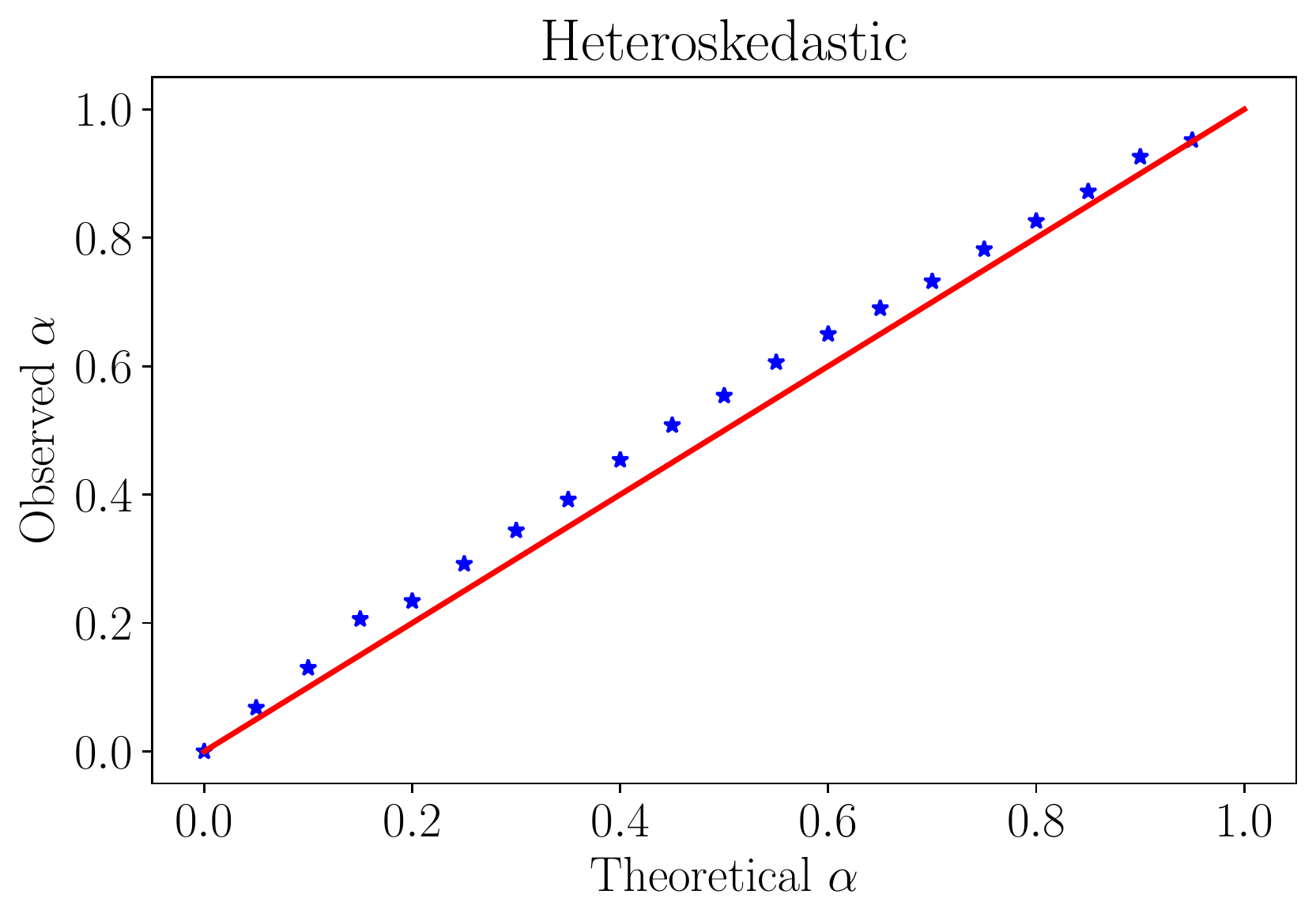}
         %\caption{$n=500$}  
     \end{subfigure}
     \hfill
     \begin{subfigure}[b]{0.48\textwidth}
         \centering
         \includegraphics[width=\textwidth]{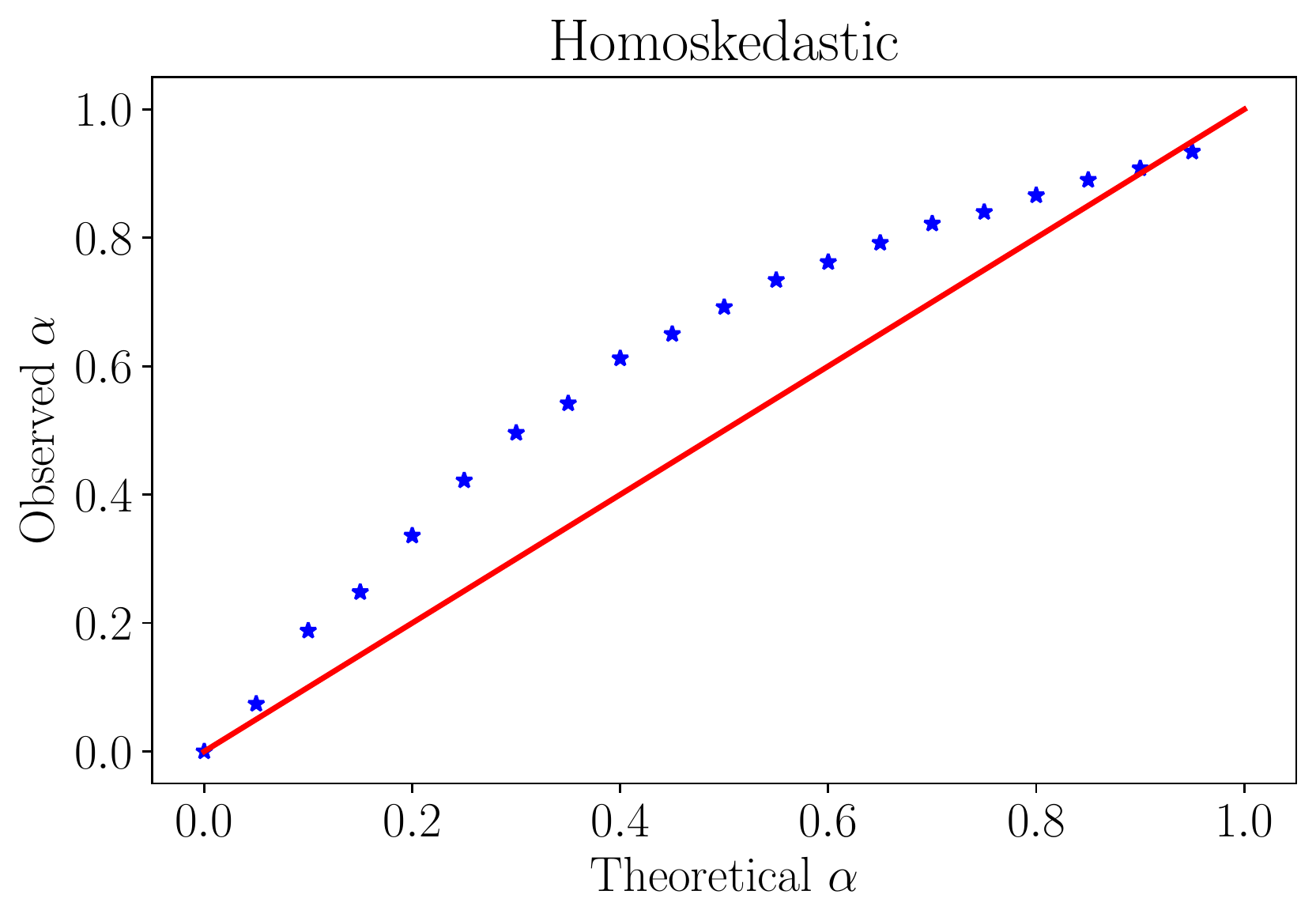}
         %\caption{$n=500$} 
     \end{subfigure}
     \caption{\label{fig:coverage} Observed proportion of the data that lies in the $\alpha$-theoretical confidence intervals with respect to their theoretical proportion for both heteroskedastic and homoskedastic Gaussian processes with a learning sample of $n=500$ nonlinear mechanical simulations.}
\end{figure}

   Moreover, we also provide a graphical tool proposed in \cite{MarrelIoossChabridon2021} which consists in evaluating the proportion of data that lies in the $\alpha$-theoretical confidence interval obtained with heteroskedastic and homoskedastic Gaussian process surrogates. Several values $\alpha \in [0,1]$ of the prediction interval level are chosen and the theoretical level of the prediction interval is compared to the empirical proportion of the data that belongs actually to this prediction interval. The empirical coverage probabilities are also estimated by leave-one-out on the learning sample of $n=500$ nonlinear mechanical simulations. By definition, the more the points are close to the identity line, the better the quality of the kriging surrogate is. Figure \ref{fig:coverage} gives the results for heteroskedastic and homoskedastic Gaussian process surrogates. We can remark that the empirical coverage probabilities with the heteroskedastic surrogate are closer to the identity line than for the homoskedastic surrogate. This can be explained by the flexibility of the variance provided by the heteroskedasticity which allows better adaptation to the distribution of the data than with a fixed value for the variance. 
   \\
   
   Finally, the observations made in \cy{Figure \ref{fig: predicted observed}} and Table \ref{tab: q2 values loo} indicate that the homoskedastic and heteroskedastic surrogates perform similarly in term of predictivity. Moreover, the heteroskedastic GP surrogate is better than the homoskedastic one to approximate the overall distribution of the data as shown with the coverage probabilities illustrated in Figure \ref{fig:coverage}. Thus, regarding the performance metrics used in this article, the heteroskedastic model is preferred to the \cy{homoskedastic} one. However, in order to validate and benchmark the methodology proposed in this paper, the two surrogate models will be used to propagate the epistemic uncertainties tainting the mechanical parameters to the seismic fragility curve and for global sensitivity indices estimation.

   \subsection{Estimation of the seismic fragility curves}
  
    \begin{figure}[!ht]
     \centering
     \begin{subfigure}[b]{0.48\textwidth}
         \centering
         \includegraphics[width=\textwidth]{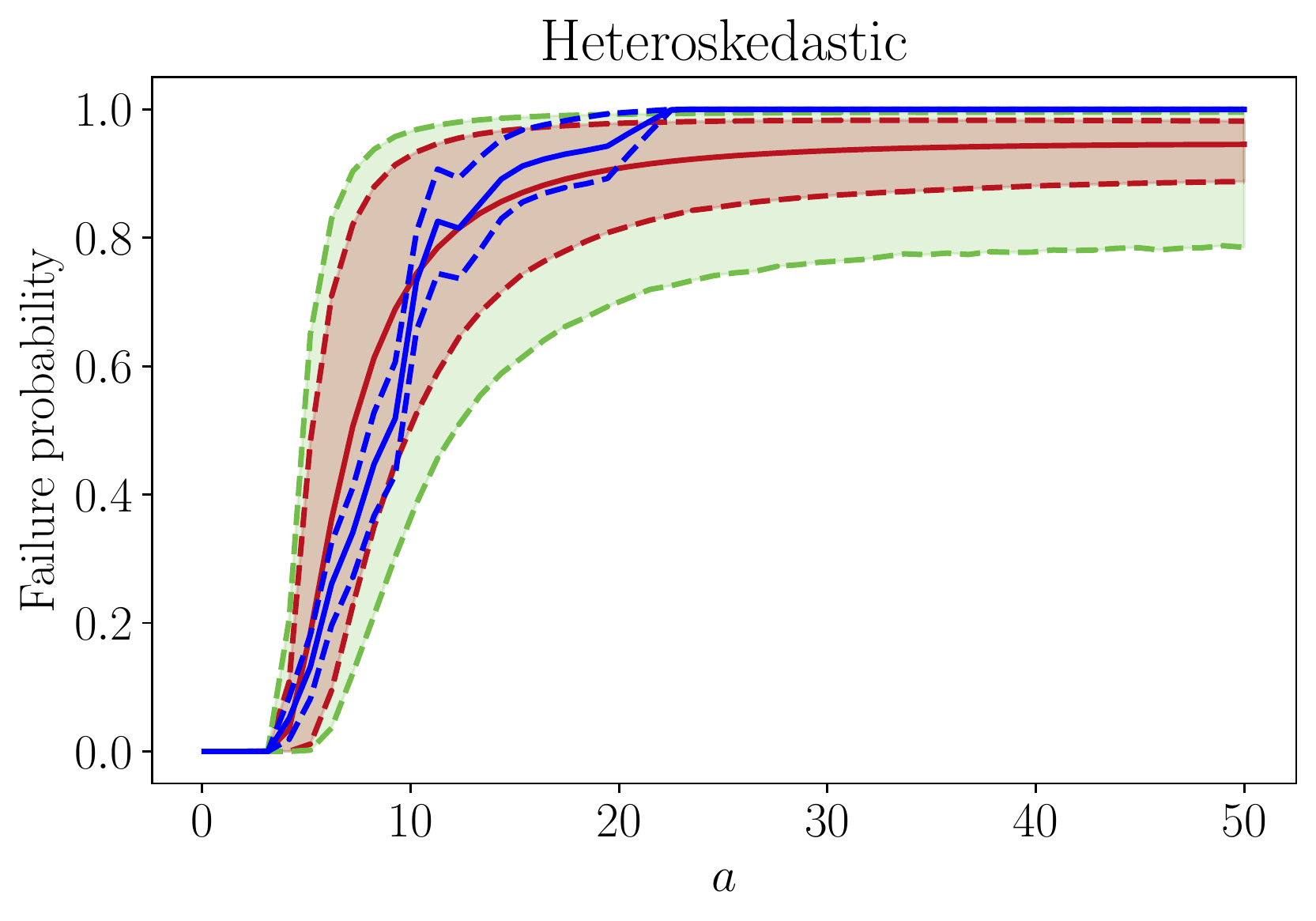}
         \caption{$C=0.5^{\circ}, \ n=200$}         
     \end{subfigure}
     \hfill
     \begin{subfigure}[b]{0.48\textwidth}
         \centering
         \includegraphics[width=\textwidth]{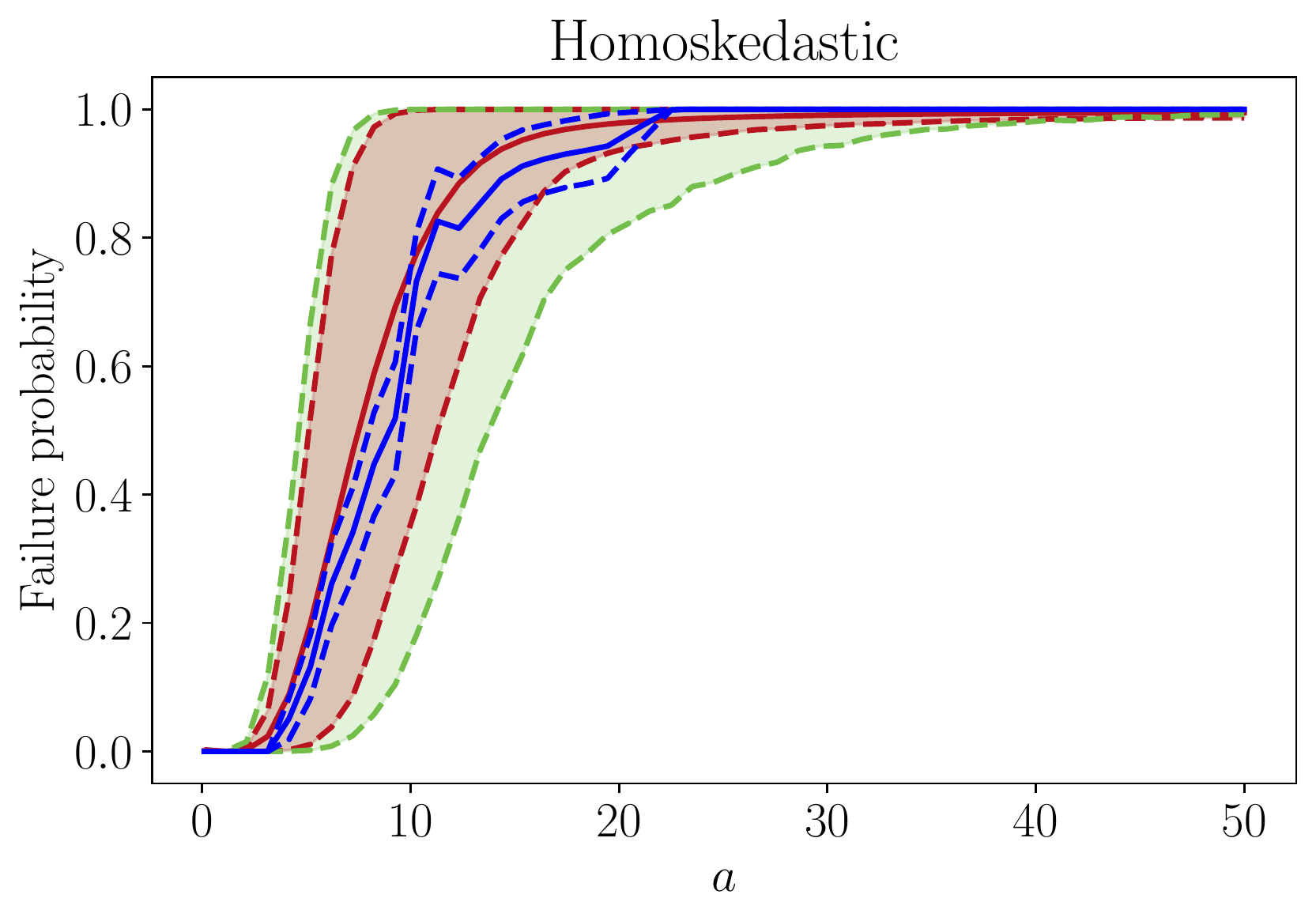}
         \caption{$C=0.5^{\circ}, \ n=200$}  
     \end{subfigure}
     \begin{subfigure}[b]{0.48\textwidth}
         \centering
         \includegraphics[width=\textwidth]{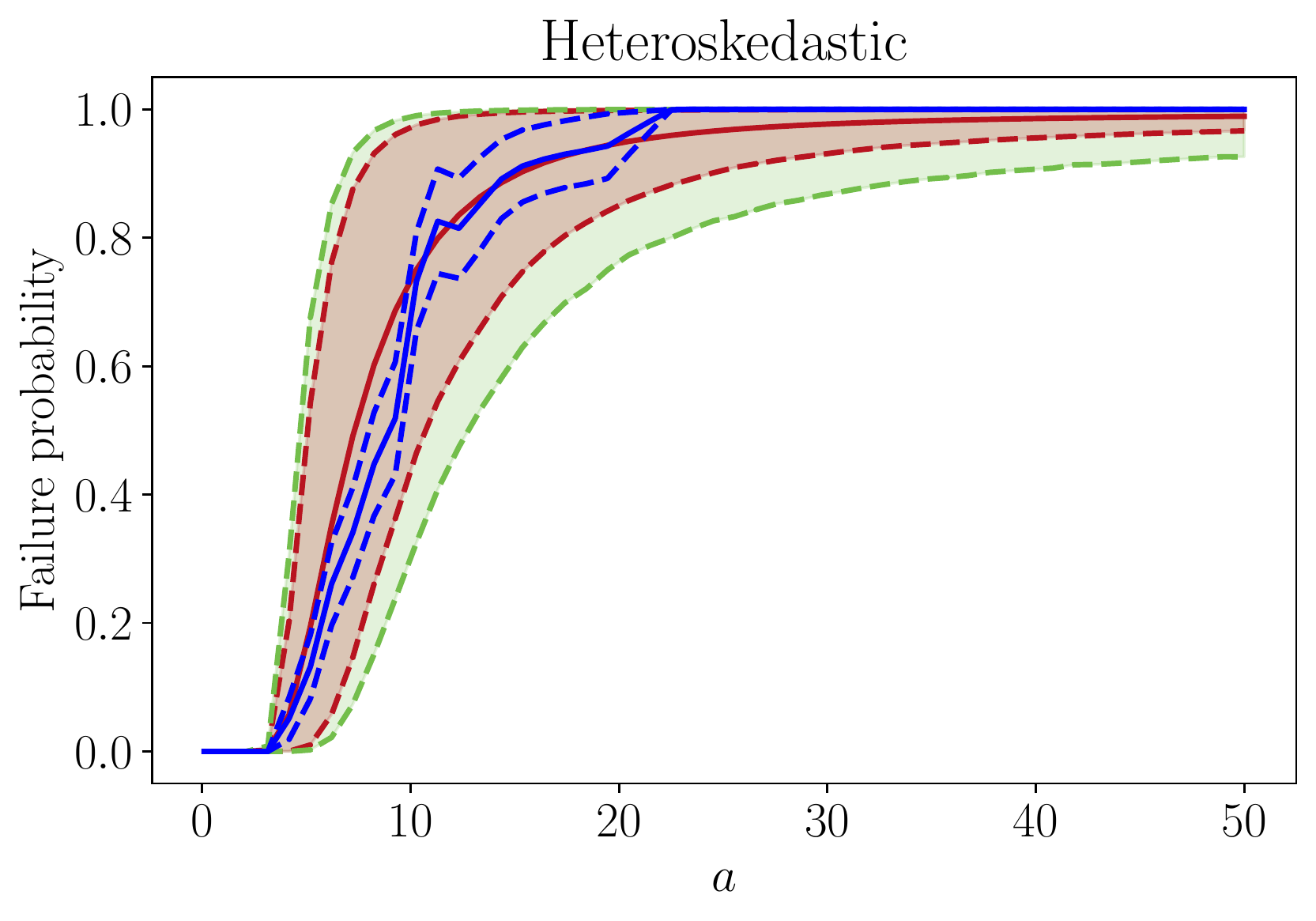}
         \caption{$C=0.5^{\circ}, \ n=500$}  
     \end{subfigure}
     \hfill
     \begin{subfigure}[b]{0.48\textwidth}
         \centering
         \includegraphics[width=\textwidth]{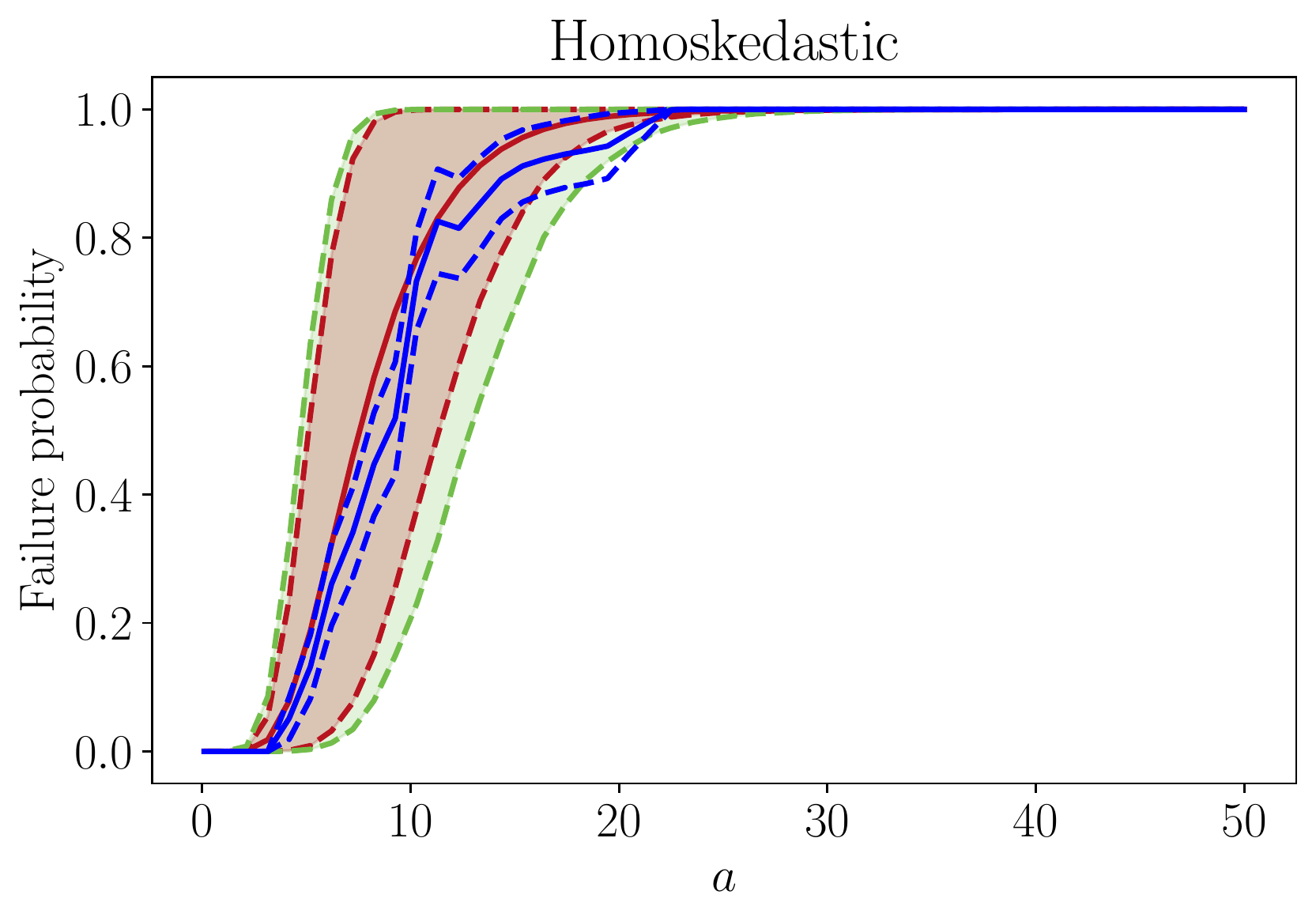}
         \caption{$C=0.5^{\circ}, \ n=500$} 
     \end{subfigure}
      \caption{\label{fig:frag UQ 0.5} Uncertainty propagation of the epistemic uncertainties on the seismic fragility curves with a failure elbow out-of-plane rotation angle $C = 0.5^{\circ}$.}
\end{figure}

\begin{figure}[!ht]
 \centering
 \begin{subfigure}[b]{0.48\textwidth}
         \centering
         \includegraphics[width=\textwidth]{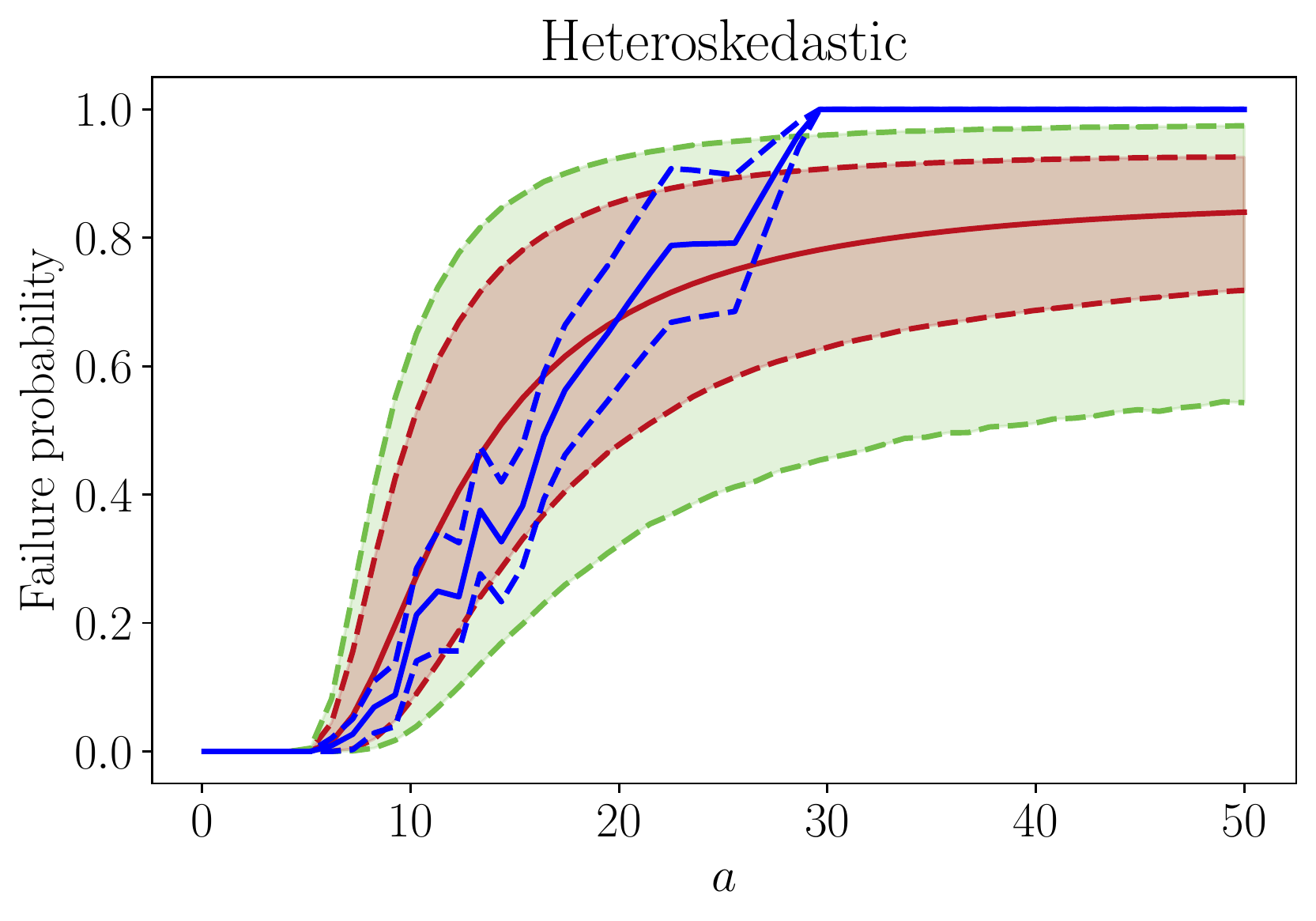}
         \caption{$C=1^{\circ}, \ n=200$}         
     \end{subfigure}
     \hfill
     \begin{subfigure}[b]{0.48\textwidth}
         \centering
         \includegraphics[width=\textwidth]{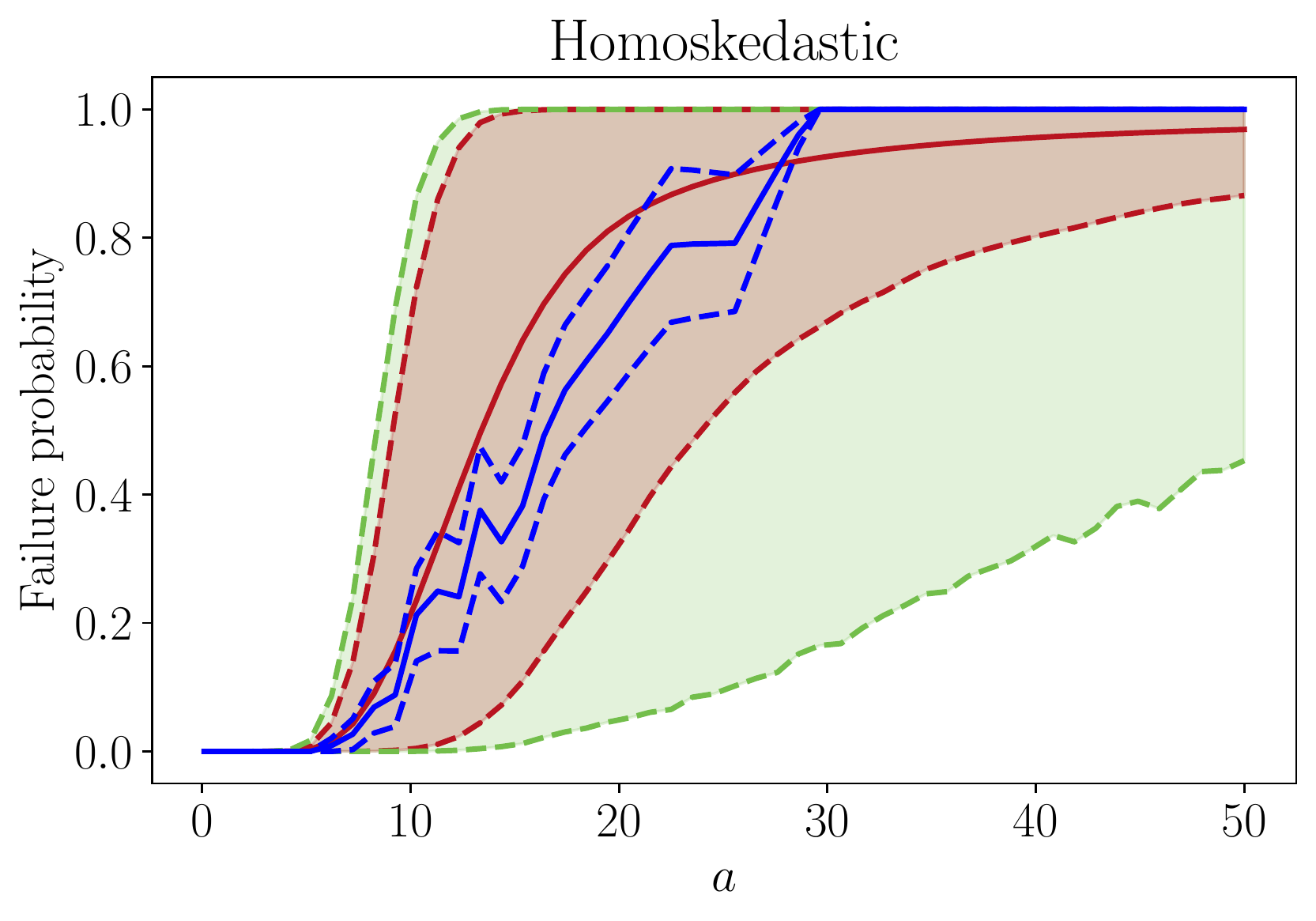}
         \caption{$C=1^{\circ}, \ n=200$}  
     \end{subfigure}
     \begin{subfigure}[b]{0.48\textwidth}
         \centering
         \includegraphics[width=\textwidth]{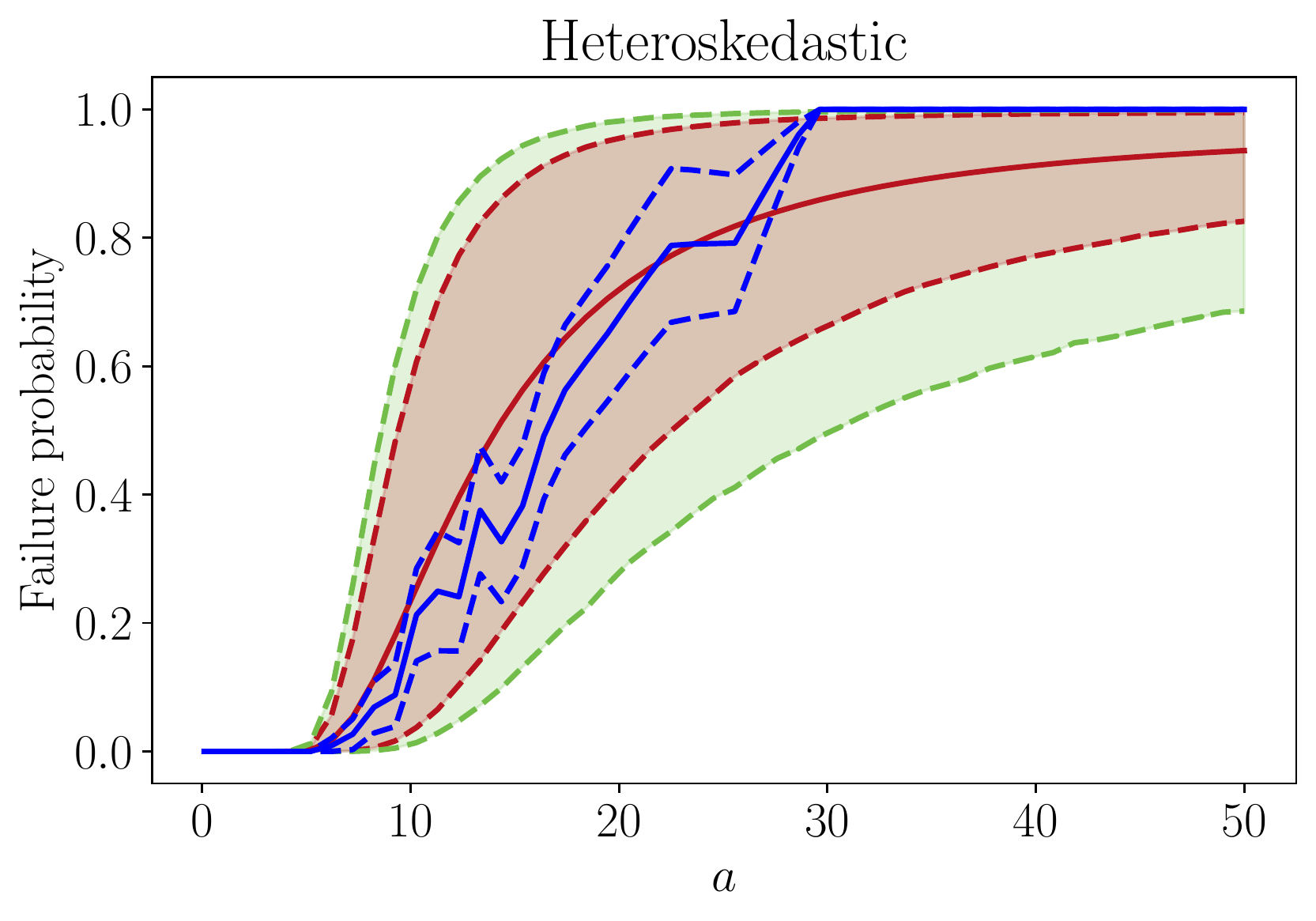}
         \caption{$C=1^{\circ}, \ n=500$}  
     \end{subfigure}
     \hfill
     \begin{subfigure}[b]{0.48\textwidth}
         \centering
         \includegraphics[width=\textwidth]{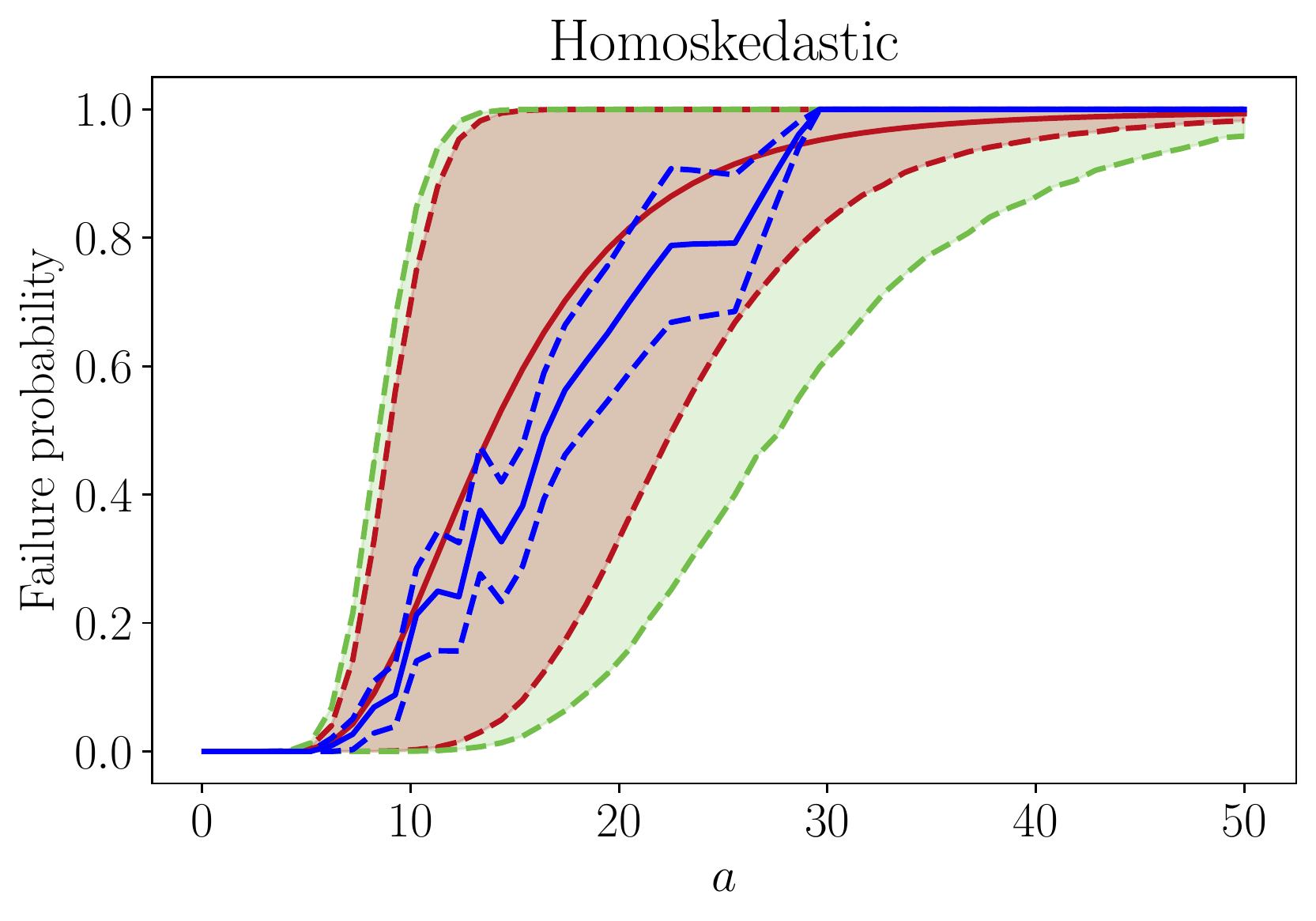}
         \caption{$C=1^{\circ}, \ n=500$} 
     \end{subfigure}
  \caption{\label{fig:frag UQ 1} Uncertainty propagation of the epistemic uncertainties on the seismic fragility curves with a failure elbow out-of-plane rotation angle $C = 1^{\circ}$.}
\end{figure}
 
The statistical quantities of interest defined in Section \ref{sec: UQ frag} are estimated empirically using a Monte-Carlo sampling $(\Xe_j)_{1 \leq j \leq m}$ of size $m=1000$. Numerical results for several training sizes $n$ and failure elbow out-of-plane rotation angles $C$ are shown in \cy{Figures} \ref{fig:frag UQ 0.5} and \ref{fig:frag UQ 1}, the red area corresponds to the area determined by the $10\%$ and $90\%$ level seismic fragility quantile curves estimated using $(\fragepigp(., \Xe_j))_{1 \leq j \leq m}$ and $(\fragepigphet(., \Xe_j))_{1 \leq j \leq m}$ for respectively the homoskedastic and heteroskedastic Gaussian \cy{processes}. The Gaussian process surrogate uncertainty is assessed by sampling $P=1000$ realizations of $\regrgp_n$ and $\gphetreal_n$ for each value $\Xe_j, \ 1 \leq j \leq m$, the bi-level seismic fragility quantile curves with $\gamma_G = \gamma_{\Xe} = 0.9$ and $\gamma_G = \gamma_{\Xe} = 0.1$ are shown in dashed green and they are estimated empirically from the datasets $(\fragepireggp_p(., \Xe_j))_{\substack{1 \leq p \leq P, \\ 1 \leq j \leq m}}$ and $(\fragepireggphet_p(., \Xe_j))_{\substack{1 \leq p \leq P, \\ 1 \leq j \leq m}}$ respectively for the homoskedastic and heteroskedastic Gaussian processes. The solid blue line corresponds to a nonparametric fragility curve estimation of the mean fragility curve using K-means clustering and binned Monte-Carlo \cite{Trevlopoulos2019} on a large dataset of $2000$ \cy{FE} simulations of the piping system (the dashed blue lines determine the $90\%$-level confidence intervals whose half-width is equal to 1.3 times the standard deviation of the empirical exceeding probability estimator in each cluster). 
We can notice that the interquantile range is larger for the homoskedastic Gaussian process than the heteroskedastic Gaussian process for small training datasets ($n=200$) and both failure elbow out-of-plane rotation angles ($C=0.5^{\circ}$ and $C=1^{\circ}$). This tends to demonstrate that the heteroskedastic surrogate fits better the conditional distribution of the log-EDP. The next section deals with the sensitivity analysis of the piping system.

\subsection{Global Sensitivity Analysis of the piping system using the Gaussian process surrogates}

\begin{figure}
    \centering
    \includegraphics[width=.7\columnwidth]{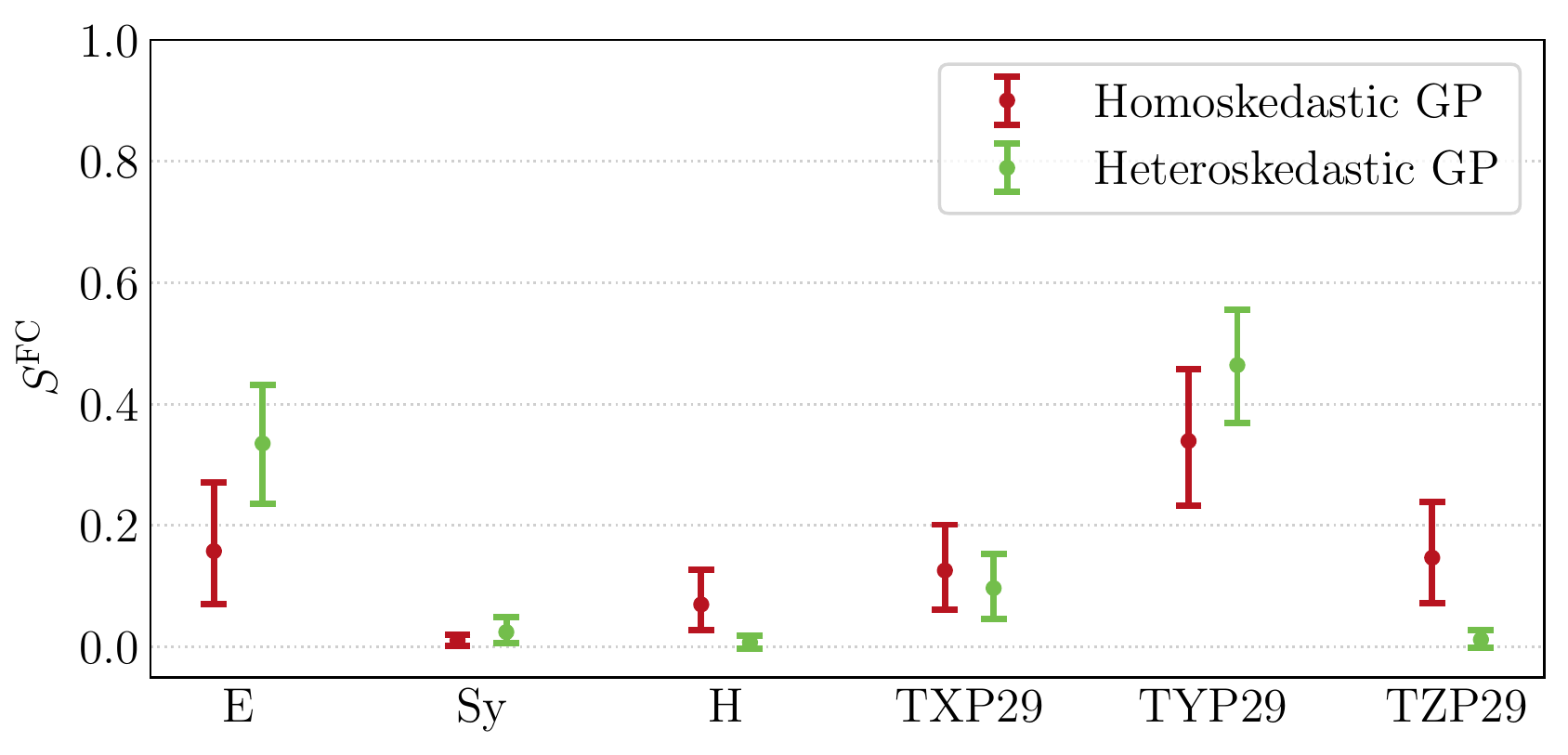}
    \caption{First-order aggregated Sobol indices for a failure rotation angle $C=1^{\circ}$ estimated with the heteroskedastic and homoskedastic GP surrogates.}
    \label{fig: sobol first ASG}
\end{figure}

\begin{figure}
    \centering
    \includegraphics[width=.7\columnwidth]{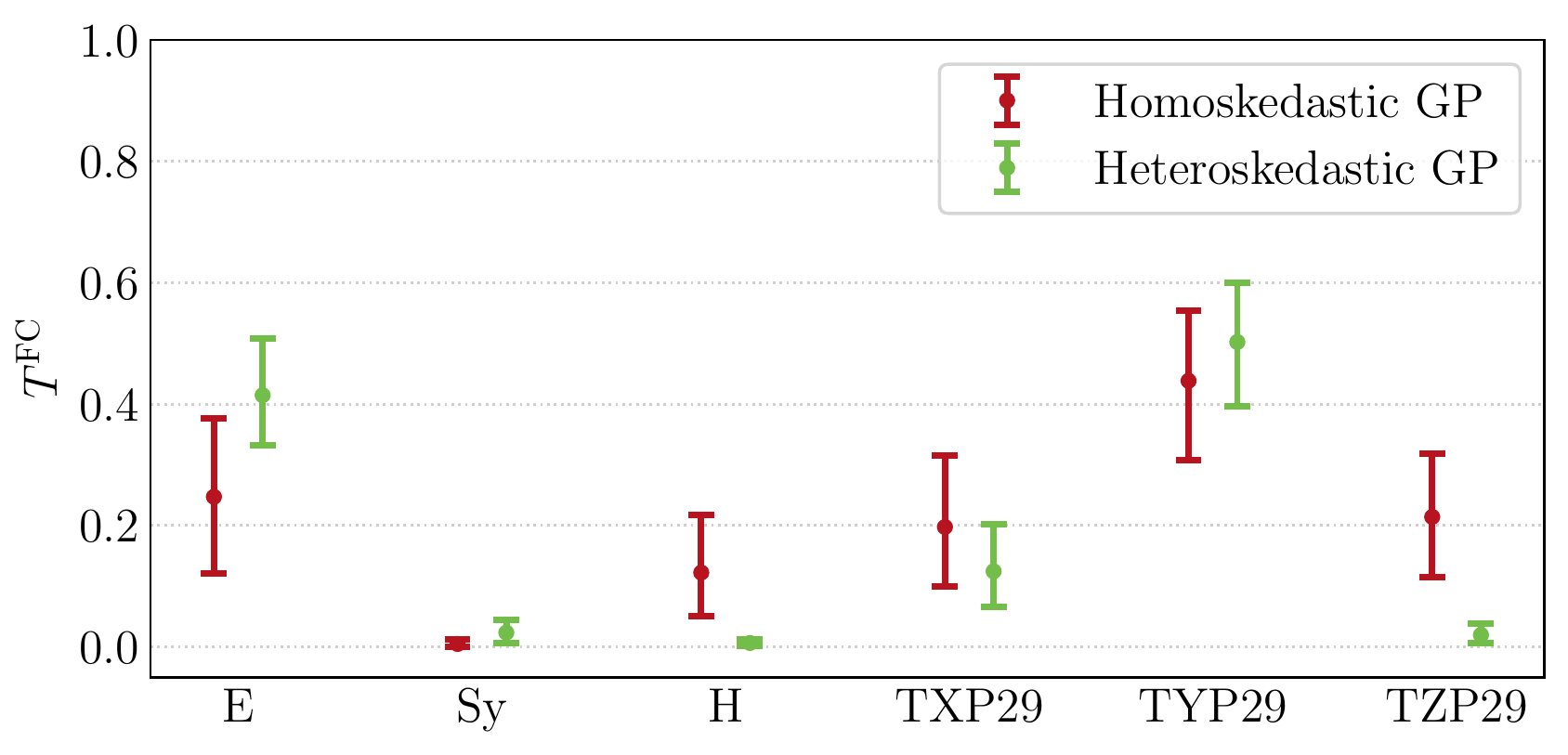}
    \caption{Total-order aggregated Sobol indices for a failure rotation angle $C=1^{\circ}$ estimated with the heteroskedastic and homoskedastic GP surrogates.}
    \label{fig: sobol tot ASG}
\end{figure}

Now we perform the estimation of the aggregated Sobol indices and the $\beta^k$ indices of the seismic fragility curves of the piping system using the methodology described in Section \ref{sec: GSA method}. A training dataset of $n=500$ simulations and a Monte-Carlo design of size $m=20000$ have been sampled in order to perform the pick-freeze estimation of the aggregated Sobol indices. $P=200$ realizations of the GP surrogate and $B=150$ bootstrap redraws have been carried out to assess the uncertainty of the aggregated Sobol indices both in terms of metamodeling and Monte-Carlo uncertainty. For the failure elbow out-of-plane rotation angle $C=0.5^{\circ}$ we compute the $L^2$ distance between fragility curves on the interval $a \in [0.1, 25]$ in order to focus on the transition area between small and high probabilities of failure. Figures \ref{fig: sobol first ASG} and \ref{fig: sobol tot ASG} provide the results for the estimation of both first-order and total-order aggregated Sobol indices for $C=1^{\circ}$ using the homoskedastic and heteroskedastic Gaussian process surrogates. 

Tables \ref{tab: std sobol hom} and \ref{tab: std sobol het} gather the numerical values of the standard deviations of the first and total order aggregated Sobol indices due to the metamodel uncertainty and the Monte-Carlo estimation uncertainty for respectively the homoskedastic and heteroskedastic Gaussian process surrogate models. Note that the standard deviation due to the Monte-Carlo estimation uncertainty is approximately ten times smaller than the one coming form the metamodel. Since increasing the sample size $n$ is more computationally expensive than increasing the Monte-Carlo sample size $m$ due to the mechanical FE computer model, it is possible to choose $m$ such that the Monte-Carlo estimation uncertainty is negligible with respect to the Gaussian process surrogate model uncertainty. The interquantile ranges represented in \cy{Figures} \ref{fig: sobol first ASG} and \ref{fig: sobol tot ASG} thus mostly come from the Gaussian process uncertainty.   

\begin{figure}
    \centering
    \includegraphics[width=.7\columnwidth]{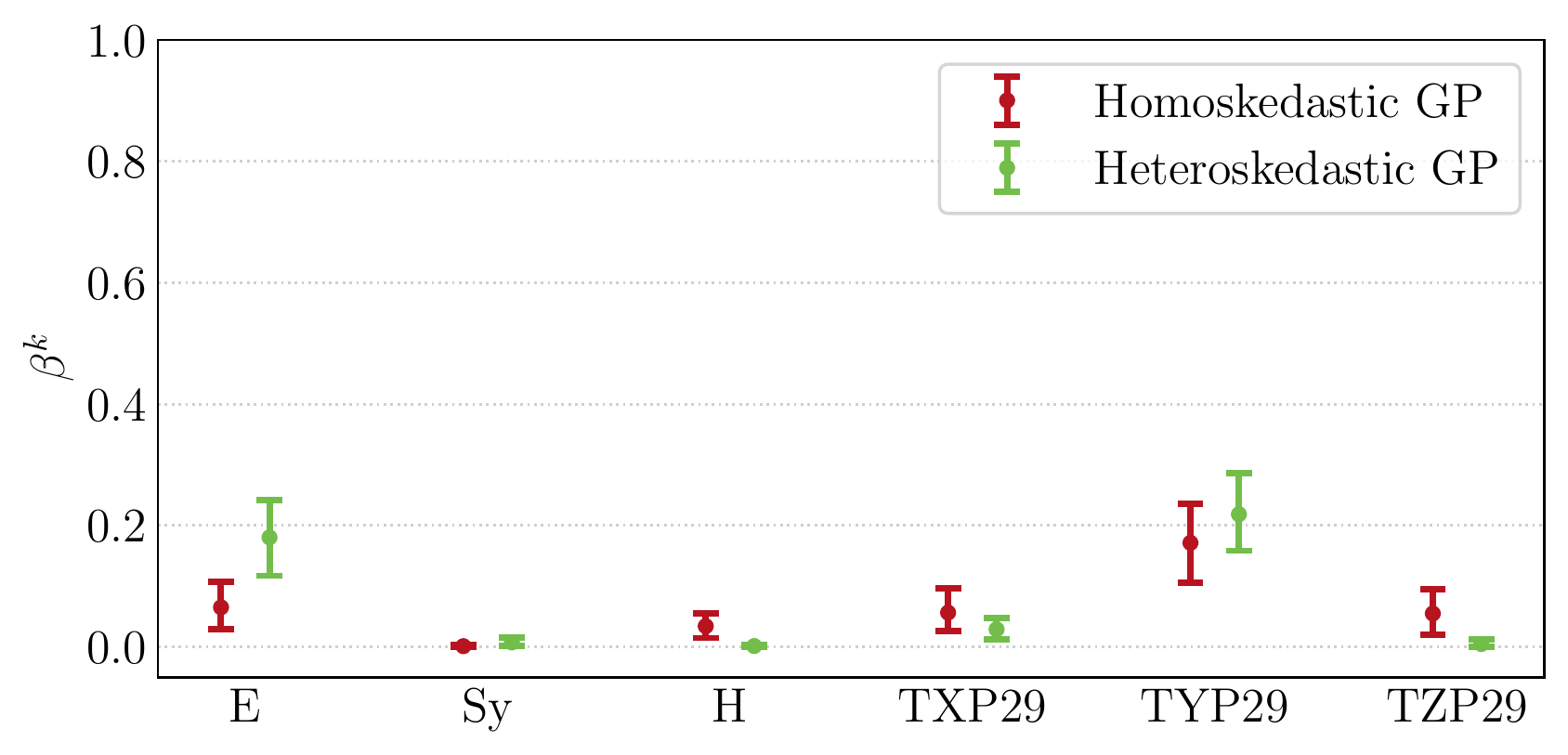}
    \caption{First-order MMD-based indices for a failure rotation angle $C=1^{\circ}$ estimated with the heteroskedastic and homoskedastic GP surrogates.}
    \label{fig: betak first ASG}
\end{figure}

\begin{figure}
    \centering
    \includegraphics[width=.7\columnwidth]{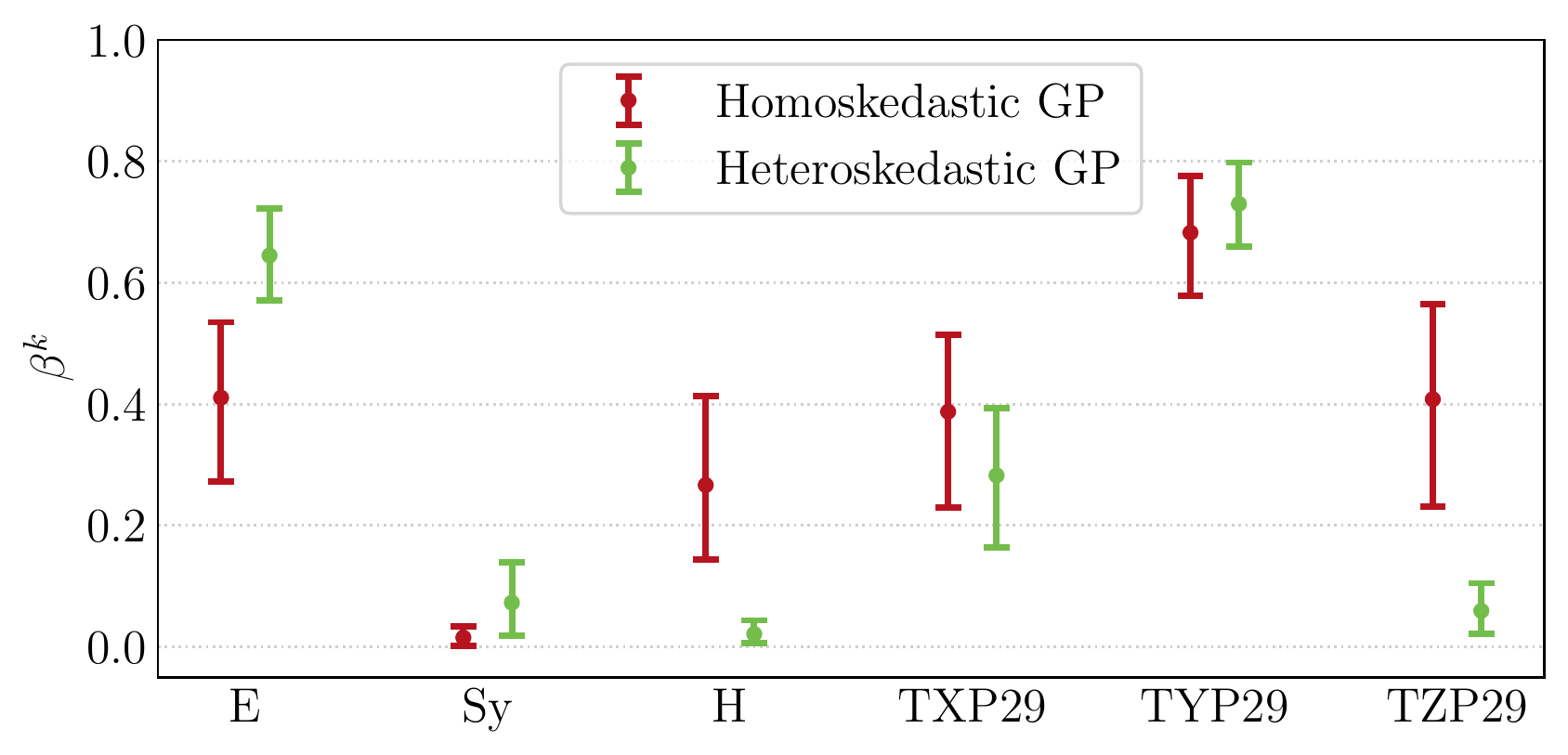}
    \caption{Total-order MMD-based indices for a failure rotation angle $C=1^{\circ}$ estimated with the heteroskedastic and homoskedastic GP surrogates.}
    \label{fig: betak tot ASG}
\end{figure}

Remark that the parameters E, TPX29 and TPY29 are the most influential on the seismic fragility curve. Indeed, the modal properties of the piping system essentially drive its dynamic behavior and hence its robustness under seismic loading. \cy{The variable TYP29 corresponds to the stiffness of the clamped end in the Y direction (i.e. the direction of the permanent loading due to the piping system's weight). What can explain why TPY29} is the most influential mechanical parameter of the piping system is the coupling of the \cy{main} eigenmodes between the X direction (i.e. the direction of the seismic load) and the Y direction. The influence of variable TYP29 is more clearly detected by the heteroskedastic Gaussian process surrogate, however the two metamodels raise the same ranking of mechanical parameters in terms of aggregated Sobol indices.
The results of the estimation of the $\beta^k$ sensitivity indices are shown in \cy{Figures} \ref{fig: betak first ASG} and \ref{fig: betak tot ASG}. We use the same parameters $n, \ P, \ B$ as for the estimation of the aggregated Sobol indices. However, we choose $m=15000$ for the Monte-Carlo design used for the $\beta^{k}$ indices pick-freeze estimator. First remark that the ranking of inputs is the same as for the one obtained with aggregated Sobol indices. However we can remark that the $\beta^{k}$ indices of the total order indices have larger values than the first order indices whereas the aggregated Sobol indices of first and total order have very close values. This means that the aggregated Sobol indices fail to detect interactions between input parameters. On the contrary, because the $\beta^k$ indices take into account the overall probability distribution of the fragility curves conditional to the input parameters, it is not surprising to detect more clearly interactions between inputs.

Tables \ref{tab: std betak hom} and \ref{tab: std betak het} gather the numerical values of the standard deviation of the $\beta^k$ indices apportioned to the Monte-Carlo estimation uncertainty and to the Gaussian process surrogate model uncertainty, respectively for the homoskedastic and heteroskedastic Gaussian process surrogate models. Similarly as for the aggregated Sobol indices, most of the uncertainty on the $\beta^k$ indices comes from the metamodel uncertainty. The interquantile ranges shown in Figures \ref{fig: betak first ASG} and \ref{fig: betak tot ASG} mostly come from the uncertainty induced by Gaussian process metamodeling.

Note that the $\beta^{k}$ indices suffer from a lack of interpretability compared to the aggregated Sobol indices: The choice of the kernel (or the choice of the lengthscale $\ell$ in the case of the Gaussian kernel) is still an open question for sensitivity analysis purposes \cite{Rabitz2022}. Similarly to the aggregated Sobol indices, the influence of TYP29 seems more clearly detected by the heteroskedastic Gaussian process surrogate than the homoskedastic one, while keeping the same ranking of influence for each mechanical parameter.   

\begin{table}[ht!]
    \centering
    \begin{tabular}{c|c|c|c|c|c|c}
        & E & Sy & H & TXP29 & TYP29 & TZP29 \\ \hline 
         $\widehat{\sigma}_{{\rm MC}_m}(\sobolestimgp)$ & $0.006$ & $0.005$ & $0.006$ & $0.006$ & $0.006$ & $0.006$ \\[2pt]
 
        $\widehat{\sigma}_{\regrgp_n}(\sobolestimgp)$ & $0.08$ & $0.007$ & $0.04$ & $0.05$ & $0.08$ & $0.07$ \\[2pt]
      
         $\widehat{\sigma}_{{\rm MC}_m}(\sobolestimgp)$ / $\widehat{\sigma}_{\regrgp_n}(\sobolestimgp)$ & $0.08$ & $0.8$ & $0.15$ & $0.10$ & $0.07$ & $0.1$ \\[2pt]
       
        $\widehat{\sigma}_{{\rm MC}_m}(\soboltotestimgp)$ & $0.004$ & $0.0004$ & $0.003$ & $0.003$ & $0.005$ & $0.004$ \\[2pt]
       
         $\widehat{\sigma}_{\regrgp_n}(\soboltotestimgp)$ & $0.1$ & $0.005$ & $0.07$ & $0.08$ & $0.1$ & $0.08$ \\[2pt]
        
         $\widehat{\sigma}_{{\rm MC}_m}(\soboltotestimgp)$ / $\widehat{\sigma}_{\regrgp_n}(\soboltotestimgp)$ & $0.04$ & $0.1$ & $0.05$ & $0.04$ & $0.05$ & $0.04$ \\[2pt]       
    \end{tabular}
    \caption{Numerical values of the part of variance of $\sobolestimgp$ and $\soboltotestimgp$ related to the Monte-Carlo estimation and to the homoskedastic Gaussian process metamodel uncertainty.}
    \label{tab: std sobol hom}
\end{table}

\begin{table}[ht!]
    \centering
    \begin{tabular}{c|c|c|c|c|c|c}
        & E & Sy & H & TXP29 & TYP29 & TZP29 \\ \hline
         $\widehat{\sigma}_{{\rm MC}_m}(\sobolestimgp)$ & $0.006$ & $0.007$  & $0.007$ & $0.007$ & $0.006$ & $0.007$ \\[2pt]
 
        $\widehat{\sigma}_{\regrgp_n}(\sobolestimgp)$ & $0.08$  & $0.02$ & $0.01$ & $0.04$ & $0.07$ & $0.01$\\[2pt]
      
         $\widehat{\sigma}_{{\rm MC}_m}(\sobolestimgp)$ / $\widehat{\sigma}_{\regrgp_n}(\sobolestimgp)$ & $0.08$ & $0.4$  & $0.7$ & $0.2$ & $0.08$ & $0.6$ \\[2pt]
       
        $\widehat{\sigma}_{{\rm MC}_m}(\soboltotestimgp)$ & $0.005$ & $0.001$ & $0.0007$ & $0.003$ & $0.006$ & $0.001$ \\[2pt]
       
         $\widehat{\sigma}_{\regrgp_n}(\soboltotestimgp)$ & $0.07$ & $0.02$ & $0.006$ & $0.05$ & $0.08$ & $0.01$ \\[2pt]
        
         $\widehat{\sigma}_{{\rm MC}_m}(\soboltotestimgp)$ / $\widehat{\sigma}_{\regrgp_n}(\soboltotestimgp)$ & $0.07$  & $0.08$ & $0.1$ & $0.06$ & $0.07$ & $0.08$ \\[2pt]       
    \end{tabular}
    \caption{Numerical values of the part of variance of $\sobolestimgp$ and $\soboltotestimgp$ related to the Monte-Carlo estimation and to the heteroskedastic Gaussian process metamodel uncertainty.}
    \label{tab: std sobol het}
\end{table}

\begin{table}[ht!]
    \centering
    \begin{tabular}{c|c|c|c|c|c|c}
        & E & Sy & H & TXP29 & TYP29 & TZP29 \\ \hline
         $\widehat{\sigma}_{{\rm MC}_m}(\betakestimgp_i)$ & $0.004$ & $0.0006$ & $0.003$ & $0.004$ & $0.005$ & $0.004$\\[2pt]
 
        $\widehat{\sigma}_{\regrgp_n}(\betakestimgp_i)$ & $0.03$ & $0.002$ & $0.02$ & $0.03$ & $0.05$ & $0.03$ \\[2pt]
      
         $\widehat{\sigma}_{{\rm MC}_m}(\betakestimgp_i)$ / $\widehat{\sigma}_{\regrgp_n}(\betakestimgp_i)$ & $0.13$ & $0.4$ & $0.15$ & $0.14$ & $0.11$ & $0.13$ \\[2pt]
       
        $\widehat{\sigma}_{{\rm MC}_m}(\betakestimgp_{-i})$ & $0.005$ & $0.0004$  & $0.004$ & $0.005$ & $0.006$ & $0.005$ \\[2pt]
       
         $\widehat{\sigma}_{\regrgp_n}(\betakestimgp_{-i})$ & $0.1$ & $0.02$ & $0.1$ & $0.07$ & $0.1$ & $0.04$ \\[2pt]
        
         $\widehat{\sigma}_{{\rm MC}_m}(\betakestimgp_{-i})$ / $\widehat{\sigma}_{\regrgp_n}(\betakestimgp_{-i})$ & $0.05$ & $0.02$ & $0.04$ & $0.05$  & $0.08$ & $0.04$ \\[2pt]       
    \end{tabular}
    \caption{Numerical values of the part of variance of $\betakestimgp$ of first and total order related to the Monte-Carlo estimation and to the homoskedastic Gaussian process metamodel uncertainty.}
    \label{tab: std betak hom}
\end{table}

\begin{table}[ht]
    \centering
    \begin{tabular}{c|c|c|c|c|c|c}
        & E & Sy & H & TXP29 & TYP29 & TZP29 \\ \hline
         $\widehat{\sigma}_{{\rm MC}_m}(\betakestimgp_i)$ & $0.004$ & $0.0006$ & $0.003$ & $0.004$ & $0.005$ & $0.004$ \\[2pt]
 
        $\widehat{\sigma}_{\regrgp_n}(\betakestimgp_i)$ & $0.03$ & $0.002$ & $0.02$ & $0.03$ & $0.05$ & $0.03$ \\[2pt]
      
         $\widehat{\sigma}_{{\rm MC}_m}(\betakestimgp_i)$ / $\widehat{\sigma}_{\regrgp_n}(\betakestimgp_i)$ & $0.1$ & $0.4$ & $0.2$ & $0.1$ & $0.1$ & $0.1$ \\[2pt]
       
        $\widehat{\sigma}_{{\rm MC}_m}(\betakestimgp_{-i})$ & $0.005$ & $0.0004$ & $0.004$ & $0.005$ & $0.006$ & $0.005$ \\[2pt]
       
         $\widehat{\sigma}_{\regrgp_n}(\betakestimgp_{-i})$ & $0.1$ & $0.02$ & $0.1$ & $0.1$ & $0.1$ & $0.1$ \\[2pt]
        
         $\widehat{\sigma}_{{\rm MC}_m}(\betakestimgp_{-i})$ / $\widehat{\sigma}_{\regrgp_n}(\betakestimgp_{-i})$ & $0.05$ & $0.02$ & $0.04$ & $0.05$ & $0.08$ & $0.04$ \\[2pt]       
    \end{tabular}
    \caption{Numerical values of the part of variance of $\betakestimgp$ of first and total order related to the Monte-Carlo estimation and to the heteroskedastic Gaussian process metamodel uncertainty.}
    \label{tab: std betak het}
\end{table}

%\clearpage

\section{Conclusion}
This work focused on the development of \cy{a comprehensive} uncertainty quantification methodology for seismic risk assessment, with a peculiar emphasis on the seismic fragility curve, a key quantity for assessing seismic safety of mechanical structures \cy{as part of SPRA studies}. \cy{Gaussian process regressions have been proposed to estimate seismic fragility curves, taking into account the epistemic uncertainties tainting the mechanical parameters of the structure among others. Gaussian process surrogates have indeed the main advantage to give both predictions and a quantification of the uncertainty on the predictions, which allows to assess the quality of the seismic fragility curve estimation through confidence intervals. In this sense, this methodology is in line with the spirit of the pioneering work of the 1980s on the SPRA framework, which defined a fragility curve not as a single curve (i.e. a mean curve) but as a family of fragility curves which reflects the uncertainty on the mean curve due to the lack of knowledge of the structures and their environment. In addition, two surrogate models have been} proposed, one modeling a homoskedastic noise and the other a heteroskedastic noise with a parameterized ramp function for the noise standard deviation. \cy{Then, different metrics have been proposed to assess the quality of the two surrogates both in predictivity and coverage performance, to allow the user an objective choice.} 

\cy{Additionally, the Gaussian process metamodels \cy{were used} to perform a global sensitivity analysis on the mechanical parameters of the structure, with the seismic fragility curve considered as a functional output. Global sensitivity indices such as aggregated Sobol indices and kernel indices have been proposed to know how the uncertainty on the mean seismic fragility curve is distributed according to each uncertain mechanical parameter. Uncertainty from Gaussian process surrogates was also taken into account when estimating the overall sensitivity indices.}

\cy{This methodology was finally illustrated considering an industrial test case consisting of a part a piping system of a French PWR. The uncertain parameters were the constitutive material parameters of the piping system as well as the boundary conditions.} The quality of the two surrogates was assessed both in predictivity and coverage performance, and seismic fragility curves was estimated for several failure thresholds and various sample sizes. Given the different qualitative and quantitative metrics used to assess the performance of the two metamodels to fit the conditional distribution of the log-EDP, the heteroskedastic metamodel was preferred because its predictive performance was similar to the one of the homoskedastic surrogate while raising more accurate coverage probabilities. In perspective, another model selection methodology could be carried out using for instance Bayesian Information Criterion (BIC) \cite{Schwarz1978, Bishop2006} or Aikake Information Criterion (AIC) \cite{Akaike1974}. \cy{After that, the aggregated Sobol indices were estimated with the two surrogates as well as kernel indices. The ranking of the input parameters was discussed and an interpretation for the results was proposed.}

An other main advantage of this UQ methodology is its flexibility. It can be generalized to computer codes with input parameters tainted by aleatory and epistemic uncertainties. The inputs with aleatory uncertainty are considered as penalizing inputs of the computer models as explained in \cite{MarrelIoossChabridon2021} and the quantity of interest defined in \cite[Section 6]{MarrelIoossChabridon2021} seems quite similar to seismic fragility curves with epistemic uncertainties defined in this article. Moreover, the methodology proposed in this article can be extended to other very similar quantity of interest such as such as for POD (Probability of Detection) curves estimation. 

Another natural extension of this work will be to propose an UQ methodology for the SPRA framework. Thus, the probability distribution of the seismic intensity measure can be taken into account and we will be able to perform the UQ study on the probability of failure of the structure, by marginalizing the seismic fragility curve on the probability distribution of the seismic intensity measure.                       

\bibliographystyle{unsrt}
\bibliography{main}

\end{document}

%% file: commands.tex
% Variables conventions
\newcommand{\edp}{z}

\newcommand{\im}{a}
\newcommand{\Itm}{A}
\newcommand{\xe}{\mathbf{x}}

\newcommand{\Xenum}[1]{\mathbf{X}^{({#1})}}
\newcommand{\Xenumpf}[1]{\tilde{\mathbf{X}}^{({#1})}}
\newcommand{\Xe}{\mathbf{X}}
\newcommand{\Xepf}{\tilde{\mathbf{X}}}

\newcommand{\x}{\mathbf{x}}

\newcommand{\epidim}{d}

\newcommand{\gppred}{m}
\newcommand{\siggp}{\sigma}

\newcommand{\gphetpred}{\widecheck{m}}
\newcommand{\sighetgp}{\widecheck{\sigma}}

\newcommand{\siggplatent}{s}
\newcommand{\siggphetlatent}{\widecheck{s}}

\newcommand{\gphetreal}{H}

\newcommand{\sigeps}{\sigma_{\varepsilon}}

\newcommand{\R}{\mathbb{R}}

\newcommand{\y}{y}
\newcommand{\Ygp}{Y}

\newcommand{\regr}{g}
\newcommand{\regrgp}{G}
\newcommand{\lengthscale}{\rho}
\newcommand{\nugget}{\varepsilon}

\newcommand{\dataset}{\mathcal{D}}
\newcommand{\model}{\mathcal{M}}

% Math symbols
\newcommand{\Norm}{\mathcal{N}}

\newcommand{\prob}{\mathbb{P}}
\newcommand{\esp}{\mathbb{E}}

\newcommand{\Xset}{\mathcal{X}}
\newcommand{\Aset}{\mathcal{A}}
\newcommand{\inputspace}{\mathcal{T}}
\newcommand{\probT}{\mathcal{M}_{\mathcal{T}}}
\newcommand{\mmd}{\text{MMD}}

% GSA notations
\newcommand{\sobol}{S^{FC}}
\newcommand{\soboltot}{T^{FC}}

\newcommand{\sobolestim}{\widehat{S}^{FC}}
\newcommand{\soboltotestim}{\widehat{T}^{FC}}

\newcommand{\sobolestimgp}{\tilde{S}^{FC}}
\newcommand{\soboltotestimgp}{\tilde{T}^{FC}}

\newcommand{\kernelfrag}{k_{\mathcal{F}}}

\newcommand{\betakestim}{\widehat{\beta}^k}
\newcommand{\betakestimgp}{\tilde{\beta}^k}

% Fragility curves notations
\newcommand{\fragepi}{\Psi}
\newcommand{\fragmean}{\bar{\Psi}}

\newcommand{\fragepigp}{\Psi^{(1)}}
\newcommand{\fragepireggp}{\Psi^{(2)}}

\newcommand{\fragepigphet}{\widecheck{\Psi}^{(1)}}
\newcommand{\fragepireggphet}{\widecheck{\Psi}^{(2)}}

\newcommand{\frag}{\overline{\Psi}}